\documentclass{emulateapj}  
\usepackage{amsmath}
\usepackage{amssymb}
\usepackage{amstext}  
\usepackage{amsfonts}              
\usepackage[parfill]{parskip}  
\usepackage{graphics}
\usepackage{epstopdf} 				
\usepackage{epsfig} 

\usepackage[usenames]{color}
\usepackage[hyperfootnotes=false]{hyperref} 			

\def\farcs{\hbox{$.\!\!^{\prime\prime}$}}
\def\fs{\hbox{$.\!\!^{\rm s}$}}

\received{}
\accepted{}
\slugcomment{Accepted by ApJ,  2013 January 26}

\begin{document}

\title{High-resolution radio observations of the remnant of SN 1987A at high frequencies}

\author{Giovanna Zanardo\altaffilmark{1}, L. Staveley-Smith\altaffilmark{1,4}, C. -Y. Ng\altaffilmark{2}, B. M. Gaensler\altaffilmark{3,4},\\
T. M. Potter\altaffilmark{1}, R. N. Manchester\altaffilmark{5} and A. K. Tzioumis\altaffilmark{5}}

\altaffiltext{1}{International Centre for Radio Astronomy Research (ICRAR), M468, The University of Western Australia, Crawley, WA 6009, Australia. \email{giovanna.zanardo@icrar.org}}
\altaffiltext{2}{Department of Physics, The University of Hong Kong, Pokfulam Road, Hong Kong}
\altaffiltext{3}{Sydney Institute for Astronomy (SIfA), School of Physics, The University of Sydney, NSW 2006, Australia}
\altaffiltext{4}{Australian Research Council Centre of Excellence for All-sky Astrophysics (CAASTRO)}
\altaffiltext{5}{CSIRO Astronomy and Space Science, Australia Telescope National Facility, PO Box 76, Epping, NSW 1710, Australia}

\begin{abstract}
We present new imaging observations of the remnant of Supernova (SN) 1987A at 44 GHz, performed in 2011 with the Australia Telescope Compact Array (ATCA). The $0\farcs35\times0\farcs23$ resolution of the diffraction-limited image is the highest achieved to date in high-dynamic range. We also present a new ATCA image at 18 GHz derived from 2011 observations, which is super-resolved to $0\farcs25$. The flux density is 40$\pm$2 mJy at 44 GHz and 81$\pm$6 mJy at 18 GHz. At both frequencies, the remnant exhibits a ring-like emission with two prominent lobes, and an east-west brightness asymmetry that peaks on the eastern lobe. A central feature of fainter emission appears at 44 GHz. A comparison with previous ATCA observations at 18 and 36 GHz highlights higher expansion velocities of the remnant eastern side. The 18--44 GHz spectral index is $\alpha=-0.80$ ($S_{\nu}\propto\nu^{\alpha}$). The spectral index map suggests slightly steeper values at the brightest sites on the eastern lobe, whereas flatter values are associated with the inner regions. The remnant morphology at 44 GHz generally matches the structure seen with contemporaneous X-ray and H$\alpha$ observations. Unlike the H$\alpha$ emission, both the radio and X-ray emission peaks on the eastern lobe. The regions of flatter spectral index align and partially overlap with the optically-visible ejecta. Simple free-free absorption models suggest that emission from a pulsar wind nebula or a compact source inside the remnant may now be detectable at high frequencies, or at low frequencies if there are holes in the ionised component of the ejecta.
\end{abstract}


\section{Introduction}
\label{Intro}

Radio supernovae (SNe) result from the collision between a supernova shock and the progenitor'\,\!s circumstellar medium (CSM). As the interaction between the propagating blast wave and the CSM drives the particle acceleration process, supernova remnants (SNRs) are natural laboratories for studying particle spectra and their variation in time \citep[see][for a review]{wei02}. 

Supernova 1987A in the Large Magellanic Cloud, as the only nearby  core-collapse  supernova observed to date, has provided a unique opportunity to study the evolution of the SN-CSM interaction in young radio remnants. The complex CSM distribution in the SNR is believed to have originated from a red supergiant (RSG) which has evolved into a blue supergiant (BSG) about 20,000 years before the explosion \citep{cro00}. Models of the progenitor evolution suggest that the equatorially denser, i.e.\ slower, RSG wind (\citealp{blo93, mar95}) was swept up by the faster BSG wind \citep{mor07}, thus forming high density rings. In particular, beside the central circular ring in the equatorial plane ({\it equatorial ring}, ER), observations with the {\it Hubble Space Telescope} ({\it HST}) have also revealed two outer rings that formed from the mass loss of the progenitor star,  located on either side of the equatorial plane (\citealp{jak91,pla95}), which confer to SNR 1987A a peculiar triple-ring nebula structure. 

Since the radio detection of the remnant in mid-1990  \citep{tur90}, the synchrotron emission has been generated by the shock wave propagating into the ring-shaped distribution of the CSM in the equatorial plane. Monitoring of the flux density has been regularly undertaken with the Molonglo Observatory Synthesis Telescope (MOST) at 843 MHz  and at 1.4, 2.4, 4.8 and 8.6 GHz with the Australia Telescope Compact Array (ATCA) \citep{sta92}. ATCA observations have been ongoing for $\sim$25 years (\citealp{sta93, gae97, bal01, man02, sta07, zan10}). An exponential increase of the flux density has been measured at all frequencies since day $\sim$5000 after the explosion, which is likely due to an increasing efficiency of the acceleration process of particles by the shock front \citep{zan10}.

The morphology of the non-thermal radiation emitted by relativistic electrons accelerated in the remnant, has been investigated using images at 9 GHz since 1992 \citep{sta93}, with a spatial resolution of 0\farcs5  achieved via maximum entropy super-resolution (\citealp{gae97, ng08}).  These images have provided the first insight into the marked east-west asymmetry of the radio emission. 
As receivers operating in the 12 mm band (16--26 GHz) were introduced at the ATCA in 2001, the first imaging observations at 18 GHz were undertaken in 2003 July, at an effective resolution of 0\farcs45 \citep{man05}. Very long baseline interferometry (VLBI) observations of the SNR were successful in 2007 October \citep{tin09} and 2008 November  \citep{ng11} at 1.4 and 1.7 GHz, respectively. These observations provided the first images with $\sim0\farcs1$ resolution, but with low sensitivity and dynamic range. Nevertheless, the VLBI images captured the presence of small-scale structures in bright regions \citep{ng11}. At the same time, ATCA observations at 36 GHz in 2008 April and October resulted in high-dynamic range images with an angular resolution of 0\farcs3 \citep{pot09}. 
The 36 GHz images, in combination with the 9 GHz observations at similar epochs,  were used to construct the first spectral index image of the SNR, with a resolution of $0\farcs45$. The resultant map provided the first glimpse into the spatial variations of the synchrotron spectral index across the remnant, and  hinted at the possible association between flatter spectral indices and the bright sites in both lobes. After the ATCA upgrade in mid-2009 with the Compact Array Broadband Backend (CABB) (Wilson et al. 2011), the remnant was  imaged at higher frequencies. The first resolved image at 94 GHz was produced from observations between 2011 June and August \citep{lak12}. Comparison of the 94 GHz image with data at 9 GHz yielded a low-resolution spectral index map, with a much larger region of flatter spectral index  on the eastern lobe than on the western lobe.  

This paper presents the first high-resolution image of SNR 1987A at 44 GHz and a new image at 18 GHz, both derived from ATCA observations performed in 2011. In \S~\ref{New_obs} and \S~\ref{morphology} we describe the imaging procedure and the resultant remnant morphology. In \S~\ref{expansion}, we estimate the remnant expansion rate, by comparing the 44 GHz data with the 36 GHz data from 2008, while the new 18 GHz data are compared to the 2003 datasets at 17 and 19 GHz. In \S~\ref{spectral} we assess the 18--44 GHz spectral index distribution, via different spectral mapping techniques. In \S~\ref{other_wavelenghts}, the comparison of the 44 GHz image and contemporaneous X-ray and H$\alpha$ observations, is discussed. In \S~\ref{free-free} we assess the likelihood of detecting at 44 GHz the radiation emitted by a compact source located within the inner regions of the remnant.

%
%
\begin{deluxetable*}{lccc}
\tablecaption{Observing Parameters \label{tab01}}
\tablewidth{0pt}
\tablecolumns{4}
\tablehead{\colhead{Parameter} & \colhead{18 GHz} & \multicolumn{2}{c}{44 GHz}}
\startdata
Date 		         					& 2011 Jan 26 			& 2011 Jan 24				& 2011 Nov 19\\
Day since explosion 				& 8738 				& 8736 					& 9036\\
Center frequencies\tablenotemark{$\ast$} (GHz) 	& 16.963 \& 18.964 			&  \multicolumn{2}{c}{43.026 \& 45.026}\\
No. of antennas 					& 6 					& 6						& 6\\
Array configuration 					& 6A 				& 6A 					& 1.5D\\
Total observing time(hr) 				& 9.18 				& 9.68 					& 9.82\\
Averaged rms path length ($\mu$m) 	& 274 				& 236 					& 198
\enddata
\tablenotetext{$\ast$}{CABB wideband mode on two interleaved 2.0--GHz wide frequency bands, each with 2048$\times$1--MHz channels. \\}
\end{deluxetable*}

\section{Observations}
\label{New_obs}

SNR 1987A was observed at 44 GHz with the ATCA in 2011 January and  November. The observations were performed on January 24 with the array in 6A configuration and on November 19  with the array in 1.5D configuration, with maximum baselines of 5939 m and 4439 m, respectively. The observations were taken over 2$\times$2-GHz bandwidth and centred on 43 and 45 GHz. 
The observations at 18 GHz were performed on 2011 January 26 with the ATCA in 6A configuration, in two bands, each of  2-GHz width,  centred on 17 and 19 GHz.
Atmospheric conditions were exceptional during both the 44 GHz sessions with rms of the path length fluctuations below 250 $\mu$m, thus confirming that phase stability conditions at the ATCA can be met not only in winter but also in spring and summer nights \citep{mid06}. The rms of the path length fluctuations was below 300 $\mu$m during the 18 GHz session. The parameters of the three  observing sessions are listed in Table~\ref{tab01}. 

In all observations, the standard bandpass calibrator PKS~B0637--752 was observed for 2 minutes every 90 minutes, while the phase calibrator PKS~B0530--727 was observed for 1.5 minutes every 6 minutes on the source. Uranus was used as flux density calibrator at 43 and 45 GHz. 
At 17 and 19 GHz, we used PKS~B1934-638 as primary flux density calibrator. This is tied to Mars and has a stable flux density at 12 mm-wavelength, as shown in \citet{sau03}.
The absolute flux scale of both Mars and Uranus between 1 and 50 GHz has been shown to have an uncertainty of $1-3$\% \citep{per12}.
With good atmospheric conditions and good calibration, we estimate that the flux calibration accuracy at 18 and 44 GHz is within 5\%. 
The observations were centred on RA $05^{\rm h}\;35^{\rm m}\;27\fs975$, Dec $-69^{\circ}\;16'\;11\farcs08$ (J2000), as in Potter et al. (2009). 

\begin{figure*}[htb]
\begin{minipage}[c]{160mm}
\begin{center}
\vspace{-0mm}
\advance\leftskip-5.0mm
\includegraphics[width=110mm, angle=0]{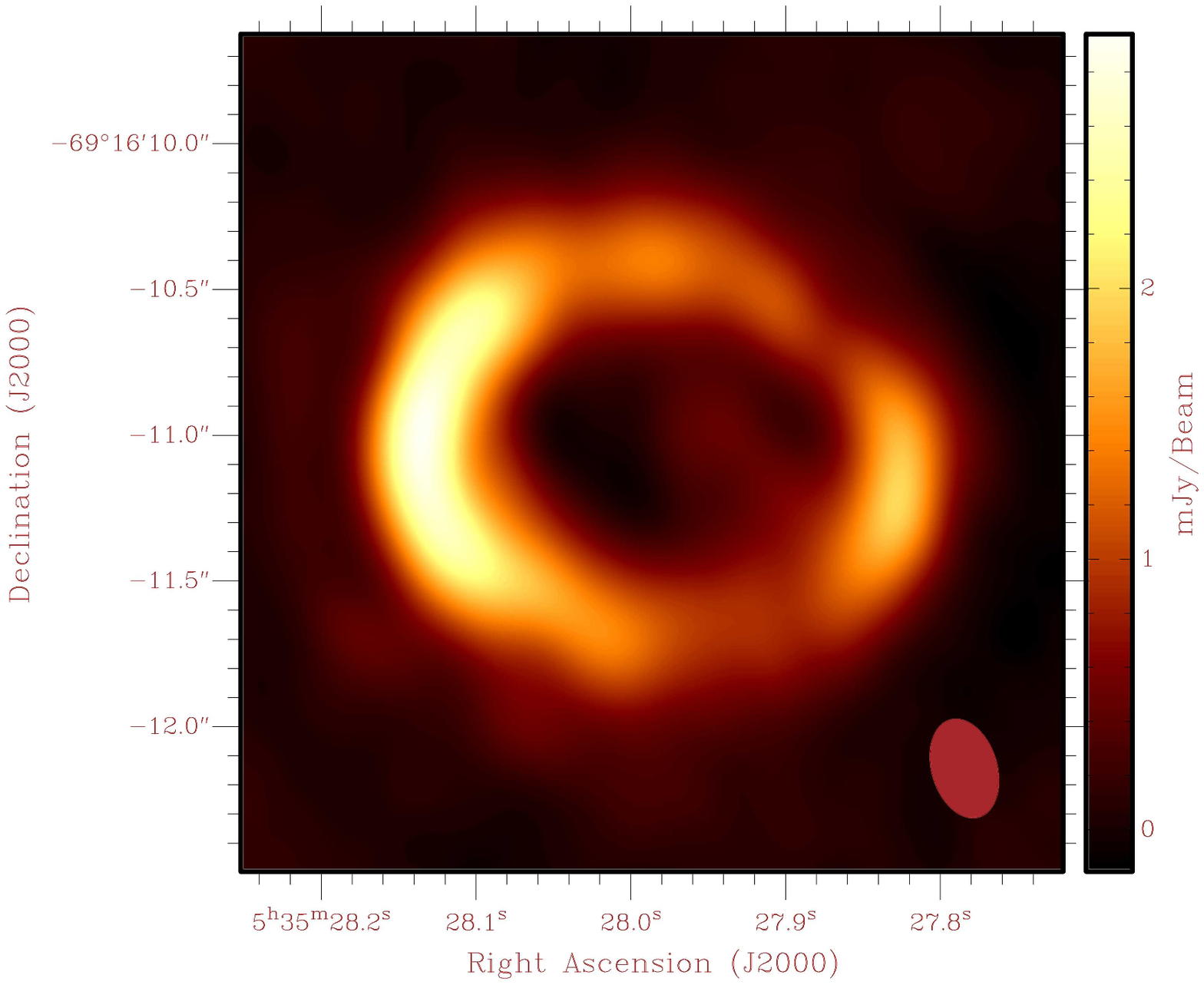}
\end{center}
\end{minipage}
\begin{minipage}[c]{160mm}
\begin{center}
\vspace{4.0mm}
\advance\leftskip+1mm
\includegraphics[width=112mm, angle=0]{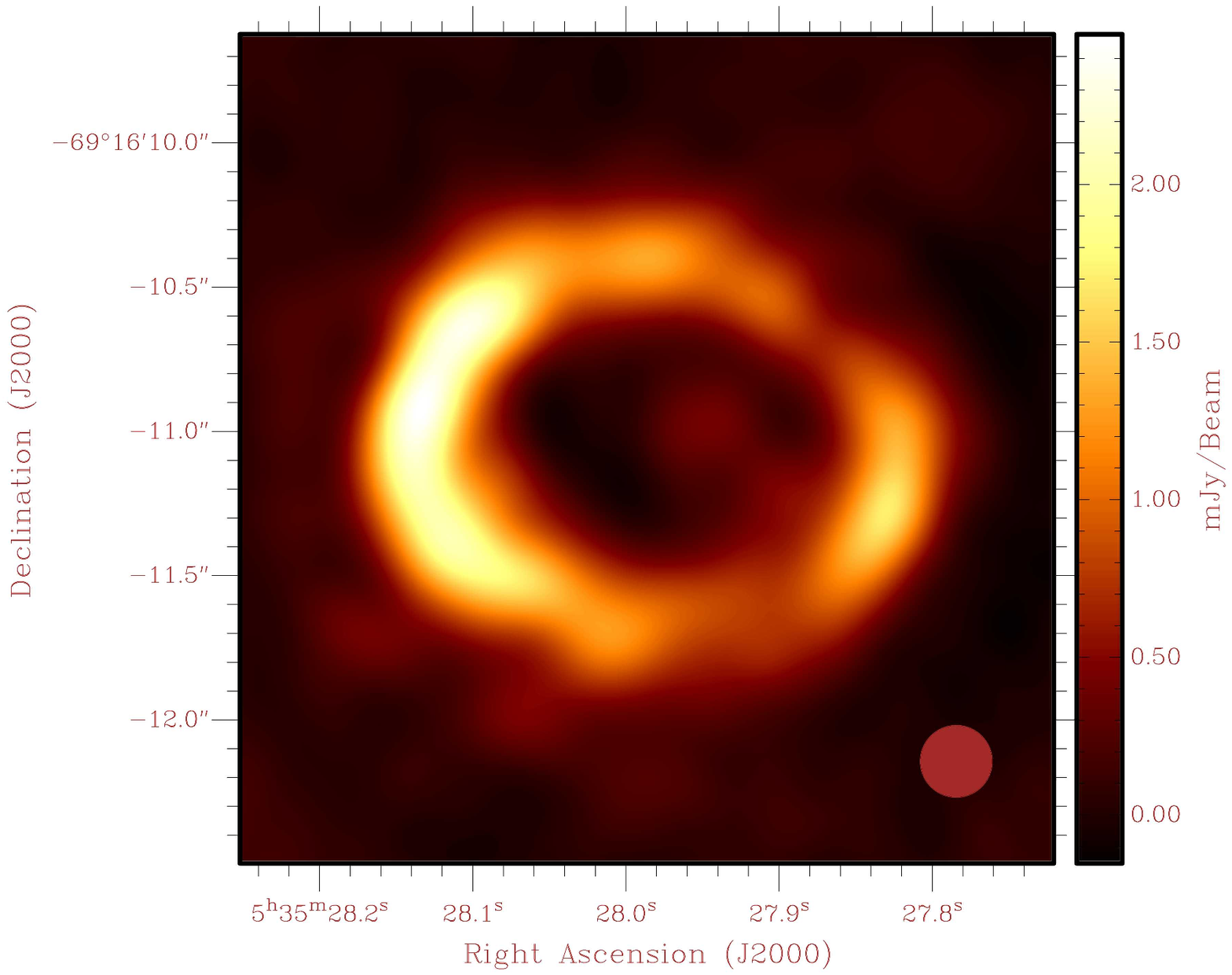}
\end{center}
\end{minipage}
\caption{{\it Top:} Diffraction-limited Stokes-I continuum image of SNR 1987A at 44 GHz made by combining observations performed with the ATCA on 2011 January 24 and 2011 November 16. The beam size is $0\farcs35\times0\farcs23$ as plotted in the lower right corner. The off-source rms. noise is $\sim$0.02 mJy beam$^{-1}$. {\it Bottom:} Slightly super-resolved 44 GHz image obtained using a $0\farcs25$ restoring circular beam (lower right corner).}
\label{fig:44_images}
\end{figure*}

The {\sc miriad}\footnote[1]{http://www.atnf.csiro.au/computing/software/miriad/} data reduction package was used to process all datasets.
For the 44 GHz observations, the task {\sc atfix} was first used to apply corrections to the system temperatures, instrumental phases and baseline lengths. 
The January and November datasets were imaged, {\sc clean}-ed  \citep{hog74} and  self-calibrated separately before being combined. In particular, the preliminary {\sc clean} model was constructed by using 400 iterations for the January dataset and 180 for the November dataset. Phase self-calibration was performed on the separate datasets and on the combined data, over a 2-minute solution interval.
For imaging, a weighting parameter of robust = 0.5  \citep{bri95} was used. Deconvolution of the combined data was performed with the maximum entropy method (MEM) \citep{gul78}. The MEM model of the combined data led to diffraction-limited image resolution of $0\farcs35\times0\farcs23$ with a moderate dynamic range ($\sim$165). The diffraction-limited image was then slightly super-resolved using a $0\farcs25$ circular beam.
Figure~\ref{fig:44_images} shows the combined Stokes-I continuum image derived from the January and November observations before and after super-resolution. The image yields an integrated flux density for the SNR of $40\pm2$ mJy at day 8886 after the explosion, where the rms error in the image and the flux density uncertainty are added in quadrature.
In Stokes-Q, U and V images, the source is not detected. The $3\sigma$ upper limits to the flux densities are 60 ${\rm \mu}$Jy beam$^{-1}$ or 2\% of the maximum Stokes-I flux density.

\begin{figure*}[htb]
\begin{minipage}[c]{160mm}
\begin{center}
\vspace{-0mm}
\advance\leftskip-6.0mm
\includegraphics[width=113mm, angle=0]{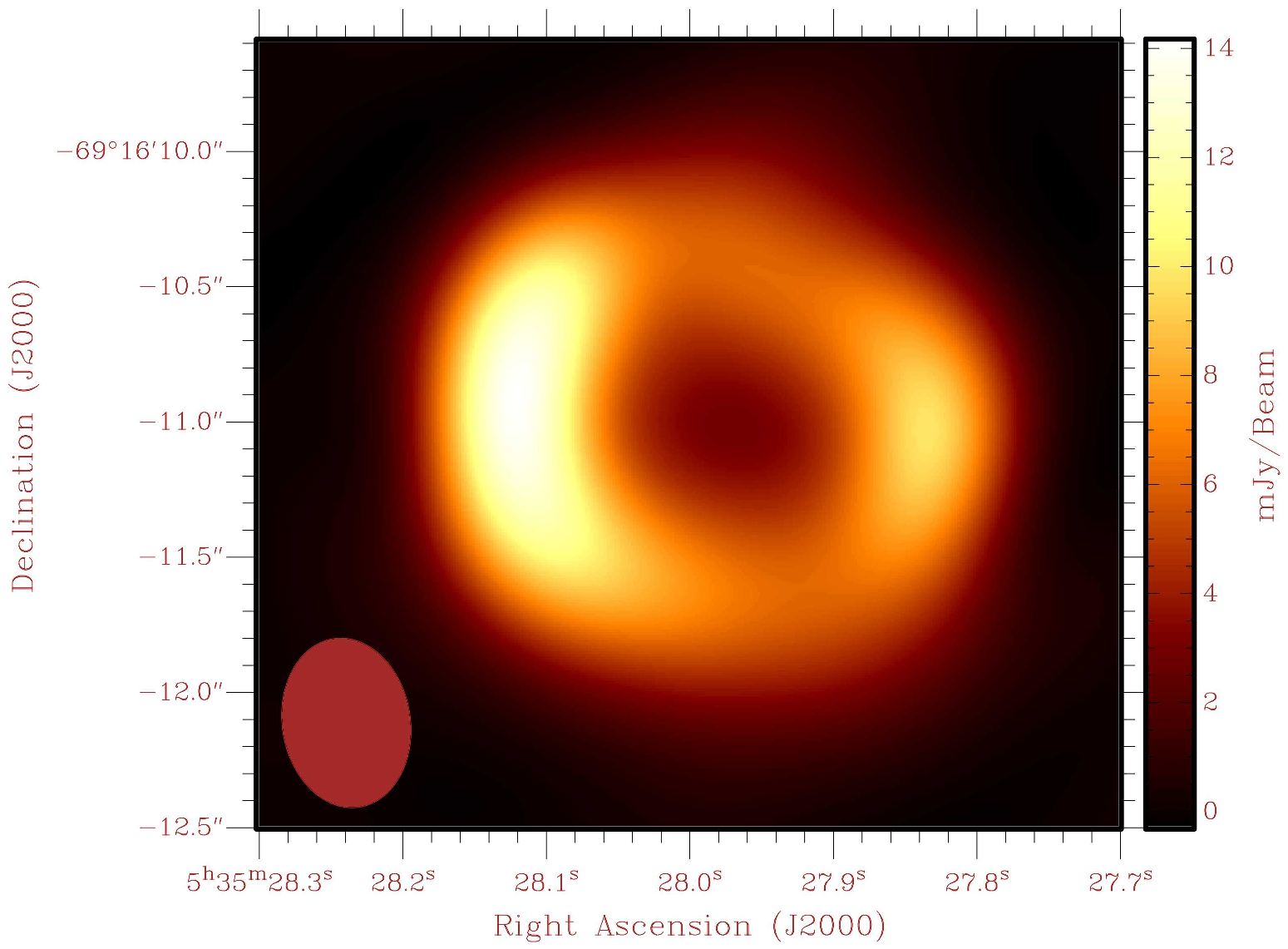}
\end{center}
\end{minipage}
\begin{minipage}[c]{160mm}
\begin{center}
\vspace{4mm}
\advance\leftskip-7.5mm
\includegraphics[width=111mm, angle=0]{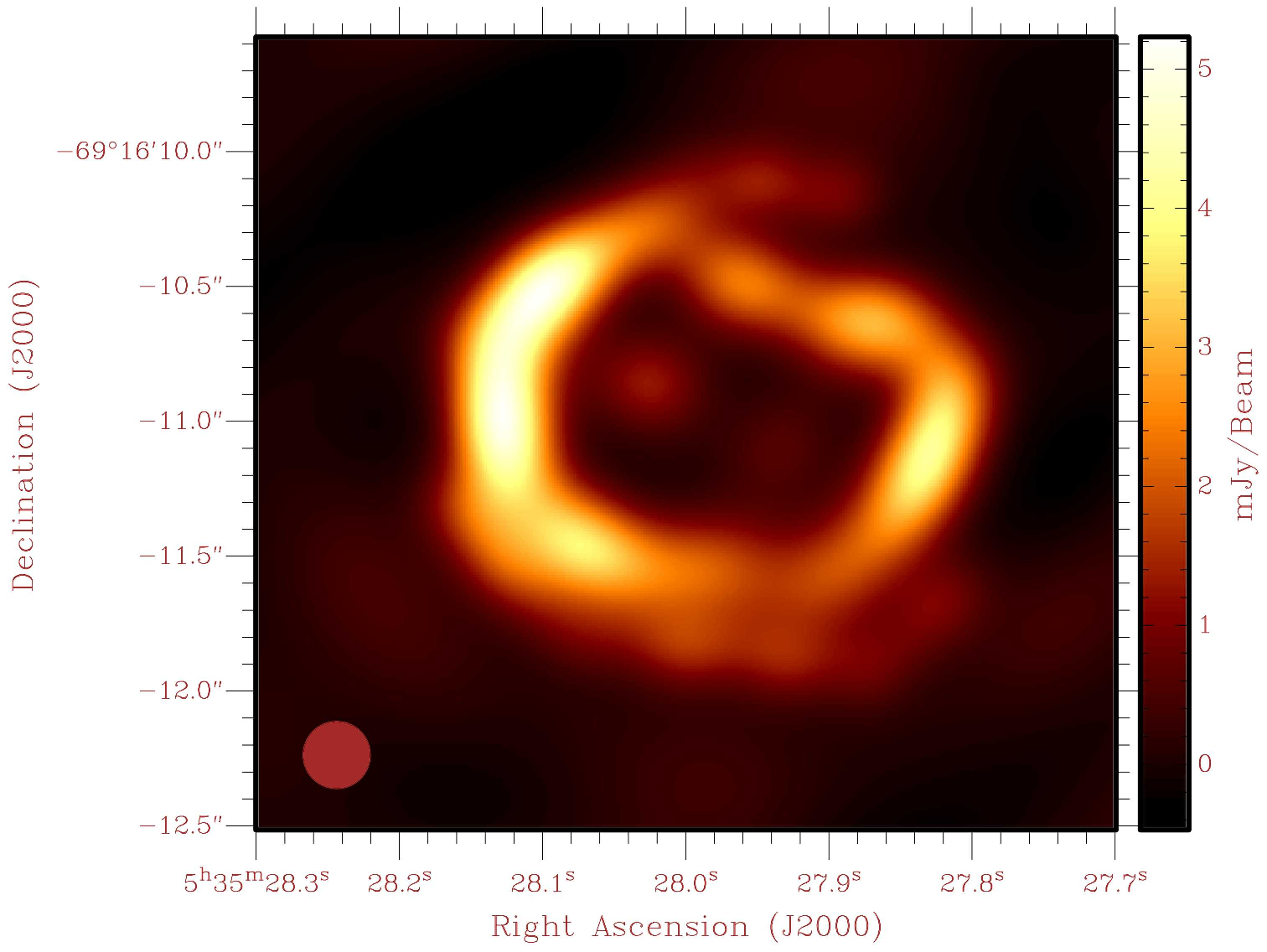}
\end{center}
\end{minipage}
\caption{{\it Top}: Diffraction-limited Stokes-I continuum image of SNR 1987A at 18 GHz from observations performed on 2011 January 26 using the ATCA. The beam size is $0\farcs63\times0\farcs47$ as plotted in the lower left corner. The off-source rms. noise is $\sim$90 $\mu$Jy beam$^{-1}$. {\it Bottom}: Super-resolved 18 GHz image obtained using a $0\farcs25$ restoring circular beam (lower left corner).}
\label{fig:18_image}
\end{figure*}

The data from the 18 GHz observations were flagged, then split into the separate observing bands until the imaging and deconvolution steps, so as to account for frequency-dependent terms in the calibration step.
Phase self-calibration was performed over a 2-minute solution interval, which gave sufficient S/N ratio.
As for the reduction procedure of the 44 GHz data, the weighting parameter robust was set at 0.5 and, after data deconvolution via MEM, the resulting image was restored with the diffraction-limited beam of half-maximum size  $0\farcs63\times0\farcs47$. Because of the high S/N,  a super-resolved image was also obtained by restoring the MEM model with a $0\farcs25$ circular beam, as \citet{man05} have shown that this degree of super-resolution gives reliable images.
The diffraction-limited and super-resolved versions of the Stokes-I continuum 18 GHz image of SNR 1987A, as derived at day 8738 since the explosion, are shown in Figure~\ref{fig:18_image}. The image gives an integrated flux density for the SNR of $81\pm6$ mJy.  The image parameters are summarised in Table~\ref{tab02}.
\vspace{0.5mm}
 
\section{Morphology of the radio emission}
\label{morphology}

\subsection{44 GHz morphology}
\label{morphology_44}

\begin{figure*}[htp]
\begin{center}
\vspace{5mm}
\advance\leftskip4mm
\includegraphics[width=97.5mm, angle=0]{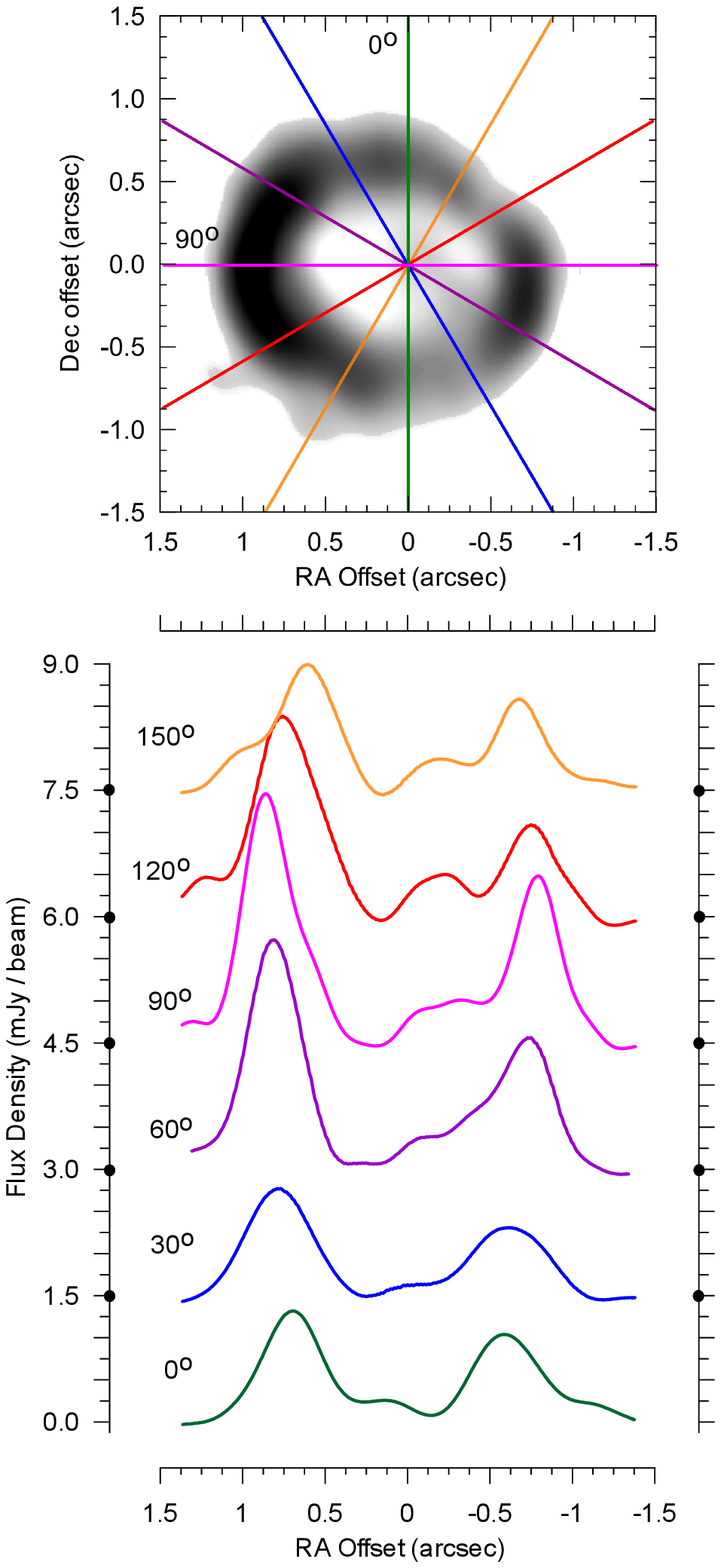}
\caption{Radial slices through the diffraction-limited 44 GHz image at 6 position angles. Black dots on vertical axes indicate the position of the zero for each slice. The offset is radial from the VLBI position of SN 1987A as determined by Reynolds et al. (1995) [RA $05^{\rm h}\;35^{\rm m}\;27\fs968$, Dec $-69^{\circ}\;16'\;11\farcs09$ (J2000)], and is positive toward east and/or north.}
\label{fig:44_profiles}
\end{center}
\end{figure*}

The emission at 44 GHz appears to be mainly distributed in an elliptical ring, with brightness peaking on the eastern lobe. The east-west asymmetry of the radio emission is  a characteristic of SNR 1987A, which has emerged since the first ATCA images \citep{gae97}, and has been monitored via the super-resolved ATCA images at 9 GHz \citep[][hereafter N08]{ng08}. This asymmetry has been confirmed in ATCA observations at 18, 36 and 94 GHz (\citealp{man05,pot09,lak12}). Slices through the 44 GHz image at six position angles are shown in Figure~\ref{fig:44_profiles}. The brightness of the eastern lobe peaks between the radial profiles at 60$^{\circ}$ and 90$^{\circ}$, while the ratio of the profile maxima corresponding to the brightness peaks of the eastern and western lobes, is $\sim$1.5. Overall, the eastern half of the image appears significantly brighter than the western one, with a ratio of $\sim$1.6 between the integrated flux densities of the two regions, i.e. east and west of the geometric centre, approximately located $\sim$75 mas east of the VLBI position of the SN  \citep{rey95} [RA $05^{\rm h}\;35^{\rm m}\;27\fs968$, Dec $-69^{\circ}\;16'\;11\farcs09$ (J2000)]. This emission ratio is higher than the  $\sim1.4$ obtained  at  9 GHz  (N08) and the $\sim1.3$ measured at  36 GHz \citep{pot09}, via fitting the images with an equatorial belt torus characterised by a west-east linear gradient of the brightness distribution (see N08).

The profiles also show a possible third peak slightly west of the centre of the remnant, which reaches maximum in the 120$^{\circ}$ slice, at $47\%$ of the brightness peak of the western lobe. The overall emission from this central region is $1.4\pm0.2$ mJy, thus making $\sim4\%$ of the total integrated flux density. 

\begin{figure}[htb]
\begin{center}
\vspace{5.0mm}
\advance\leftskip-6.75mm
\includegraphics[width=95mm,angle=0]{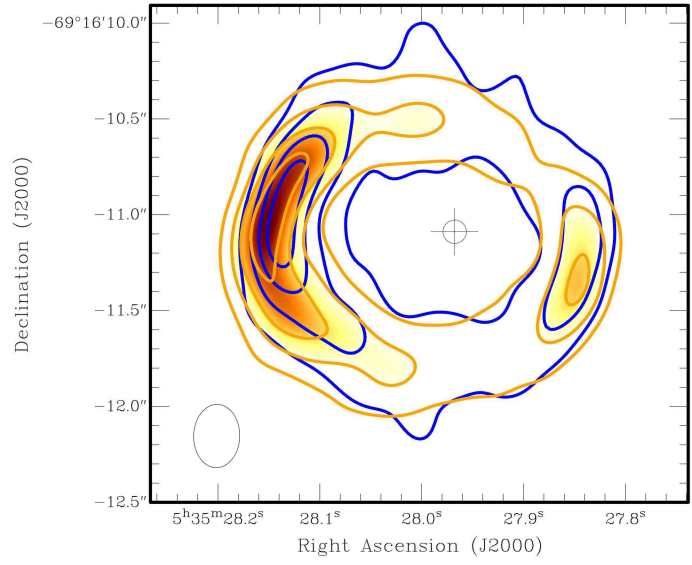}
\caption{Comparison between the 44 GHz image and that derived from ATCA observations at 36 GHz performed in 2008 \citep{pot09}. The 44 GHz image is restored with the $0\farcs33\times0\farcs24$ beam that characterises the diffraction-limited  image at 36 GHz,  which is set at PA $-1.3^{\circ}$  \citep[][Table~\ref{tab03}]{pot09}. The contours at levels $25\%-85\%$ of the peak flux density, in step of 10\%,  are in blue for the 36 GHz image and in orange for the 44 GHz image. To highlight the morphological differences of the brightest sites in the two images, the 44 GHz contours at 45\%, 65\% and 85\% of the peak flux density, are filled in yellow, orange and brown. The cross-hair symbol marks the VLBI position of SN 1987A \citep{rey95}.}
\label{fig:44_2008-2011}
\end{center}
\end{figure}

In Figure~\ref{fig:44_2008-2011}, the 44 GHz image is compared to the 2008 observations at 36 GHz (see Table~\ref{tab03}). The remnant expansion over the 4-year time frame is noticeable from the contours of the peak flux density at identical levels. 
A total integrated flux density of $23.6\pm5.2$ mJy is derived by scaling the flux density of the 36 GHz image to that associated with the remnant at 44 GHz at the same epoch, as $\alpha=-0.68$ results at day 7815 from the spectral index function that fits observations between 1.4 and 8.6 GHz \citep[see Figure 7 in][]{zan10}. This leads to a yearly flux density increase of $24\pm6\%$, which is higher than the $17\pm8\%$ per year derived at 8.6 GHz by \citet{zan10}, from monitoring observations between Year 20 and 21. 

\begin{figure*}[htb]
\begin{minipage}[c]{160mm}
\begin{center}
\vspace{-5mm}
\advance\leftskip-5.mm
\includegraphics[width=99.8mm, angle=0]{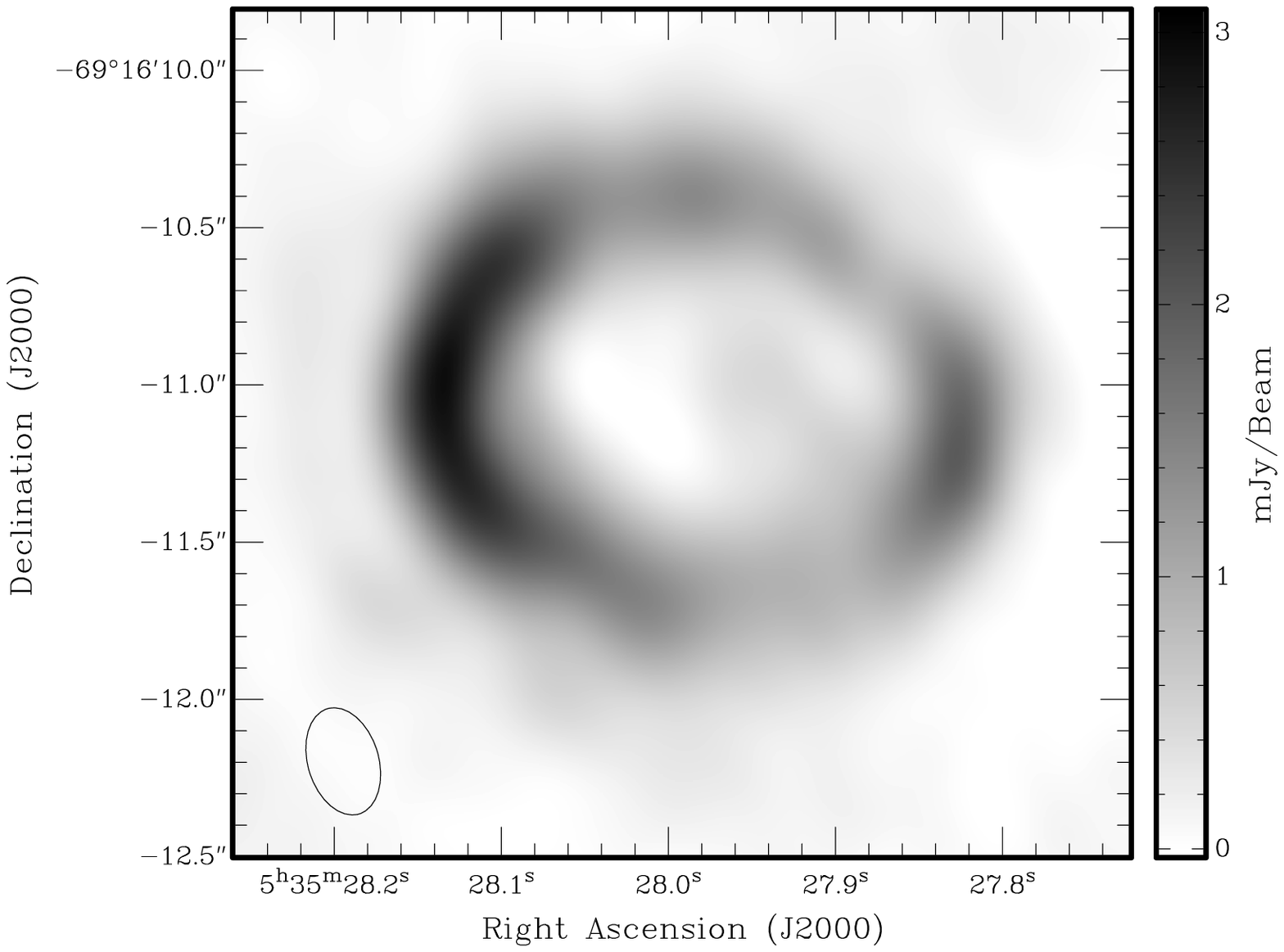}
\end{center}
\end{minipage}
\begin{minipage}[c]{160mm}
\begin{center}
\vspace{0mm}
\advance\leftskip-6.0mm
 \includegraphics[width=100mm, angle=0]{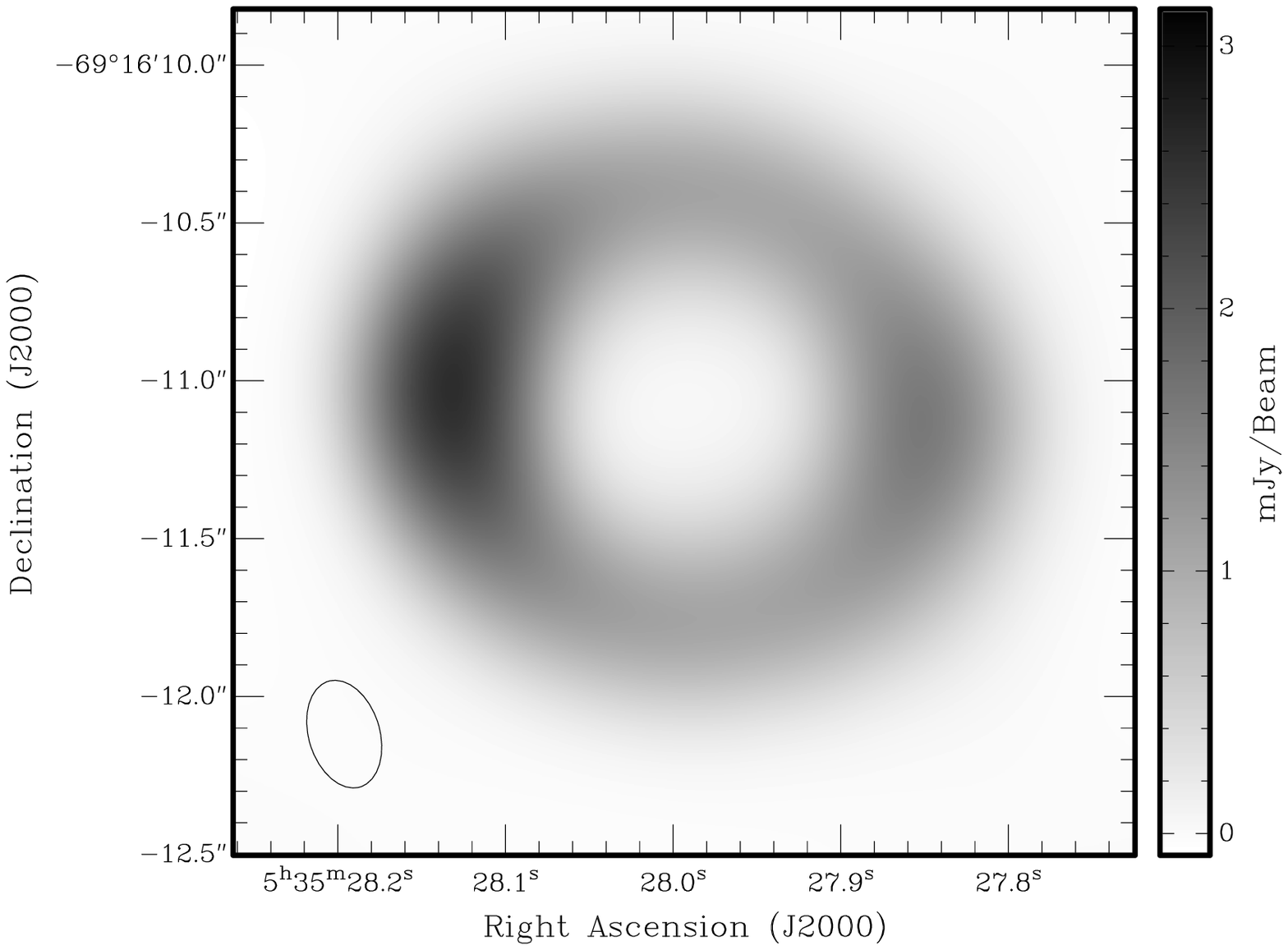}
\end{center}
\end{minipage}
\begin{minipage}[c]{160mm}
\begin{center}
\vspace{0mm}
\advance\leftskip-0.5mm
\includegraphics[width=106.5mm, angle=0]{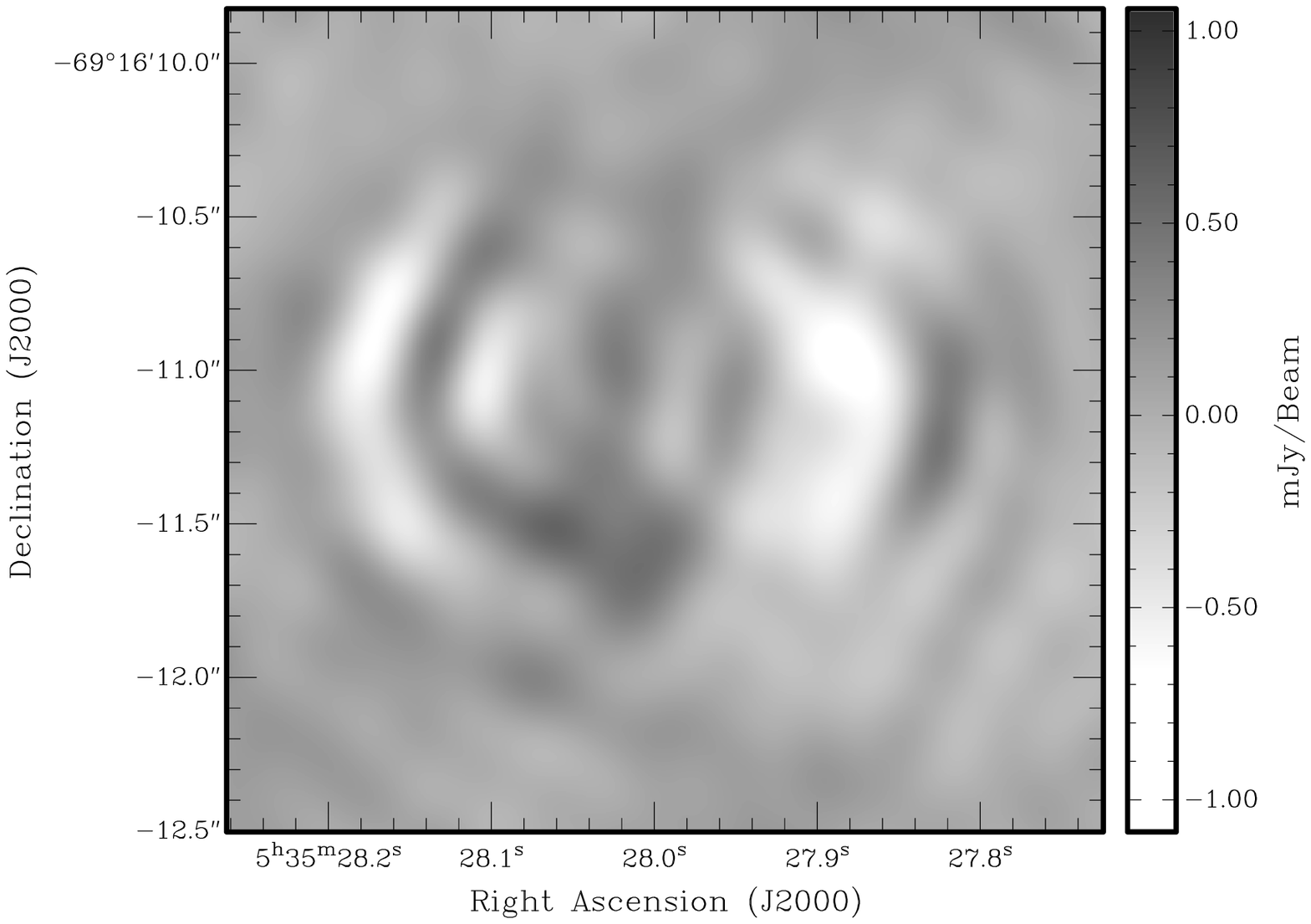}
\end{center}
\end{minipage}
\caption{Comparison of the diffraction-limited 44 GHz image ({\it top}), as derived from ATCA observations performed in 2011 January and November, with the image generated with the Fourier-domain model ({\it centre}) developed by \citet{ng08} to fit the SNR observations at 9 GHz. The model used in the comparison fits the 9 GHz observations performed in 2011 April 22 (Ng et al., in preparation), and has a radius of $0\farcs917$. To match the total flux of the observations at 44 GHz, the model flux density has been scaled using a spectral index $\alpha=-0.74$ (see Figure~\ref{fig:spectral_index}).  {\it Bottom}: Dirty map of the residual visibilities obtained by subtracting the model from the 44 GHz observations. No  deconvolution has been applied to the residual.}
\label{fig:44_3cm-model}
\end{figure*}

\begin{figure*}[!ht]
\begin{center}
\vspace{-0mm}
\includegraphics[width=110mm,angle=0]{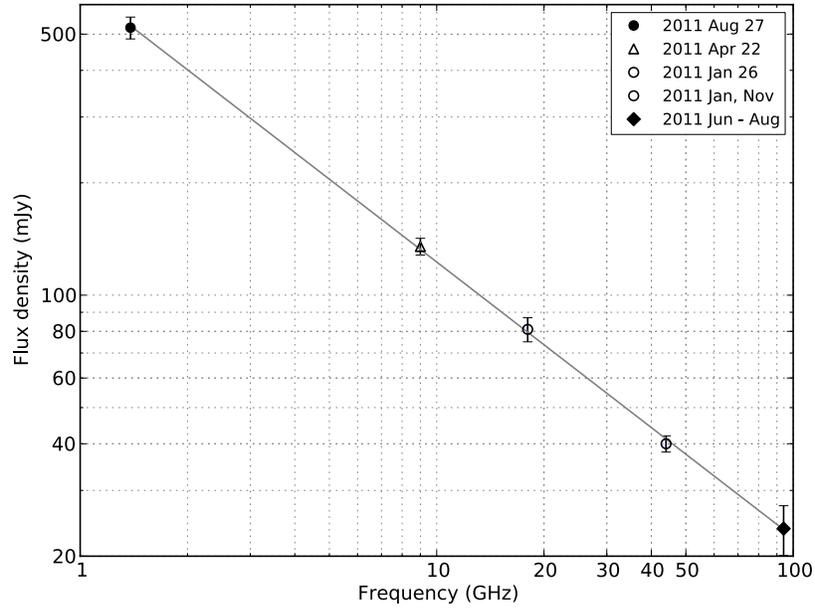}
\caption{Spectral index of SNR 1987A as determined from ATCA observations performed in 2011. In detail, the spectral index is determined from the power-law fit to the flux densities measured at  five frequencies: 1.4 GHz (Zanardo et al., in preparation) ({\it solid circle}), 9 GHz  (Ng et al., in preparation) ({\it hollow triangle}), 18 and 44 GHz (this paper) ({\it hollow circle}), and 94 GHz \citep{lak12} ({\it solid diamond}). The fit yields $\alpha=-0.74\pm0.01$.}
\label{fig:spectral_index}
\end{center}
\end{figure*}

In Figure~\ref{fig:44_3cm-model}, the 44 GHz image is compared with that derived by fitting the N08 inclined-torus model to the 9 GHz observations performed in 2011 April 22 (Ng et al., in preparation). The fitted model has a radius of $0\farcs917$, which surpasses by $\sim$67 mas that used to fit the 36 GHz image in 2008 \citep{pot09}. The flux of the 9 GHz model is scaled to that corresponding at 44 GHz, via the spectral index $\alpha=-0.74$,  which is extrapolated from contemporaneous  ATCA observations spanning from 1.4 to 94 GHz (see Figure~\ref{fig:spectral_index}). The resultant model image, as shown in Figure~\ref{fig:44_3cm-model}, has a total integrated flux density of $40\pm1$ mJy and  the restoring beam of the diffraction-limited image at 44 GHz. 
The residual between the model and the actual visibilities demonstrates that the model matches the observations along the profile at PA 90$^{\circ}$, where the ratio of the brightness peaks between the two lobes equals the $\sim1.6$ measured in the observations.  The residual visibilities peak on the eastern lobe at PA 30$^{\circ}$ and 150$^{\circ}$, since the new observations exhibit  more extended bright regions than what reproduced by the linear gradient of the  modelled flux distribution. The linear gradient assumption also constrains the ratio of the integrated flux density over the eastern and western halves of the image, to $\sim1.3$. Therefore, the N08 model, which fits well the radio remnant as seen at 9 GHz, might not be applicable to the large-scale emission morphology that emerges from the 44 GHz observations.

\subsection{18 GHz morphology}
\label{morphology_18}

In the diffraction-limited image, the emission at 18 GHz looks primarily ring-shaped, while a secondary structure seems to emerge from the super-resolved image (see Figure~\ref{fig:18_image}). Two arm-like  features appear to stem from the northern and  southern edges of the major ellipsoidal structure of the emission, peaking at PA 0$^{\circ}$, and extending towards the western side of the remnant. While these features suggest a double-ring formation within the SNR, they might more likely be noise artefacts of the super-resolution process.

\begin{figure*}[htp]
\begin{center}
\vspace{5mm}
\advance\leftskip4mm
\includegraphics[width=101.0mm, angle=0]{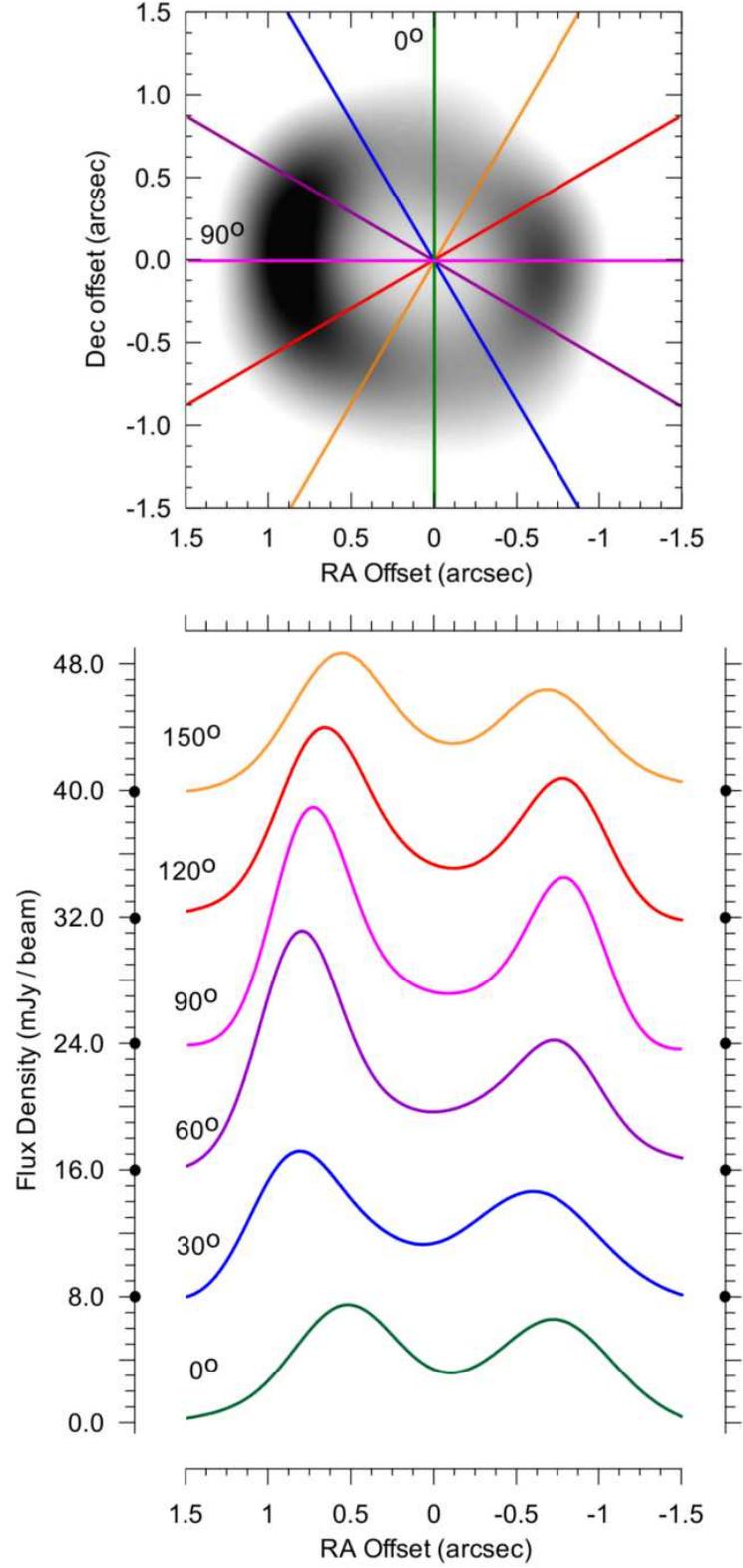}
\caption{Radial slices through the diffraction-limited 18 GHz image at 6 position angles. Black dots on vertical axes indicate the position of the zero for each slice. The offset is radial from the VLBI position of SN 1987A as determined by Reynolds et al. (1995), and is positive toward east and/or north.}
\label{fig:18_profiles}
\end{center}
\end{figure*}

In terms of remnant asymmetry, at 18 GHz  the radial profiles in Figure~\ref{fig:18_profiles} show that the asymmetry peaks on the eastern lobe between PA 60$^{\circ}$ and 90$^{\circ}$. The asymmetry ratio between the eastern and western brightness peaks is $\sim1.4$, and a similar ratio is obtained from the ratio between the flux density integrated over the eastern and western halves of the image. This value matches the asymmetry ratio derived by N08 from $1992-2008$ data at 9 GHz.

\begin{figure}[htb]
\begin{center}
\vspace{5.0mm}
\advance\leftskip-7.5mm
\includegraphics[width=95mm,angle=0]{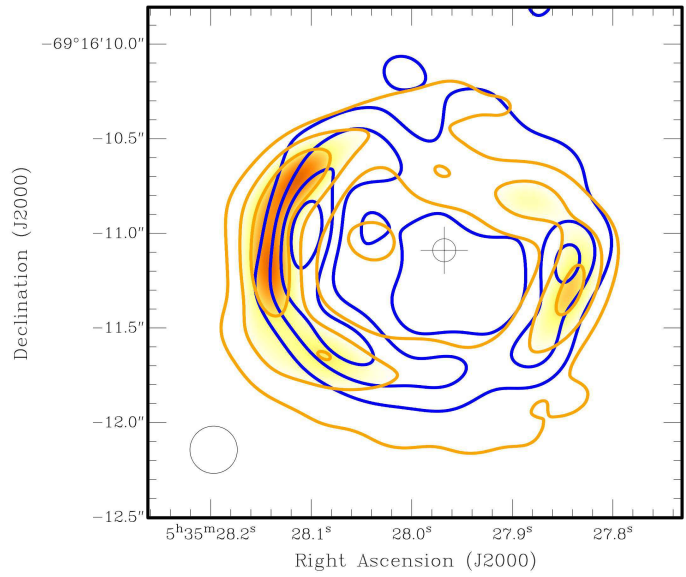}
\caption{Comparison between the new 18 GHz image and that derived from ATCA observations at 18 GHz performed in 2003 \citep{man05}. Both images are super-resolved with a $0\farcs25$ circular beam. The contours at levels $25\%-85\%$ of the peak flux density, in step of 10\%,  are in blue for the 2003 image and in orange for the new image. To highlight the morphological differences of the brightest sites in the two images and the significant expansion of the remnant since 2003, 
the contours of the new image at 45\%, 65\% and 85\% of the peak flux density, are filled in yellow, light  and dark orange. The cross-hair symbol marks the VLBI position of SN 1987A \citep{rey95}.}
\label{fig:18_2003-2011}
\end{center}
\end{figure}

In Figure~\ref{fig:18_2003-2011}, the super-resolved image is compared to that derived from previous ATCA observations at 18 GHz performed in 2003 July 31, which was also super-resolved to $0\farcs25$ \citep[][Table~\ref{tab03}]{man05}.
The morphology similarities between the 2003 and 2011 images emphasise the significant expansion of the remnant over 8 years. Slices through the 2011 image indicate that the central emission varies from 20\% to 24\% of the brightness on the eastern lobe. This figure is consistent  with the 2003 observations, where the central emission was estimated $\sim$20\% of the average ring intensity \citep{man05}.

%
%
\begin{deluxetable}{lcc}
\tablecaption{Image Parameters \label{tab02}}
\tablewidth{82mm}
\tablecolumns{3}
\tablehead{\colhead{Parameter} 					& \colhead{  44 GHz} 		& \colhead{18 GHz}}
\startdata
Integrated flux density (mJy)						& $40\pm2$				& $81\pm6$\\
Restoring beam ($''$)							& $ 0.350\times0.225$		&$0.630\times0.474$\\
Position angle 	($^{\circ}$)						& $1.674$					& $0.879$\\
Rms noise (mJy/beam)							& $0.023$					& $0.088$\\
Dynamic range									& $\sim165$				& $\sim176$
\enddata
\end{deluxetable}
\vspace{0.0mm}

%
%
\begin{deluxetable*}{lcc}
\tablecaption{Pre-CABB ATCA Images at 18 and 36 GHz \label{tab03}}
\tablewidth{0pt}
\tablecolumns{3}
\tablehead{\colhead{Parameter} 				& \colhead{18 GHz\tablenotemark{$a$}}  	& \colhead{36 GHz\tablenotemark{$b$}} }
\startdata
Observing session 		         							& 2003 July 31 					& 2008 Apr 25	 \& Oct 7,8,12\\
Day since explosion 								& 6003 						& 7815\tablenotemark{$c$}\\
Centre frequencies (GHz) 							& 17.3 \& 19.6 					& 34.9 \& 37.4\\
Flux density (mJy)  									& $27\pm4$					&$27\pm6$\\
Restoring Beam ($''$)								& $0.45\times0.39$				&$0.33\times0.24$\\
Position Angle ($^{\circ}$)								& $2^{\circ}$					&$-1.3^{\circ}$\\
Super-resolution ($''$)								& $0.25$						&$-$
\enddata
\tablenotetext{$a$}{See \citet{man05}.}
\tablenotetext{$b$}{See \citet{pot09}.}
\tablenotetext{$c$}{Average date of the combined datasets.\\}
\end{deluxetable*}

The integrated flux density of the 2003 image, as  derived from the combination of the data at 17.3 and 19.6 GHz, was $27\pm4$ mJy, which is lower than what expected from an extrapolation of the radio spectrum. This value leads to a threefold increase in flux from day 6003 to 8738, i.e.\ a yearly increase rate of  $27\pm5\%$. This increase rate  exceeds the $17\pm8\%$ per year derived at 8.6 GHz by \citet{zan10}, for data between Year 20 and 21 since the SN. 


\section{Remnant expansion}
\label{expansion}

From 1987 to 1992, i.e.\ in the early stages of the SNR, the remnant had an expansion velocity of $\sim$ 30,000 km s$^{-1}$ \citep{gae97}. When the radio emission re-emerged, this was followed by a drastic deceleration to 3000 km s$^{-1}$  \citep{gae97}.  From 1992 to 2008, the expansion has been determined from the radius of the 9 GHz images fitted via a torus model, with a resultant velocity of 4000$\pm$400 km s$^{-1}$ from $\sim$day 1800  to day 7620 (N08).

\begin{figure*}[!ht]
\begin{center}
\vspace{-0mm}
\includegraphics[width=14.0cm,angle=0]{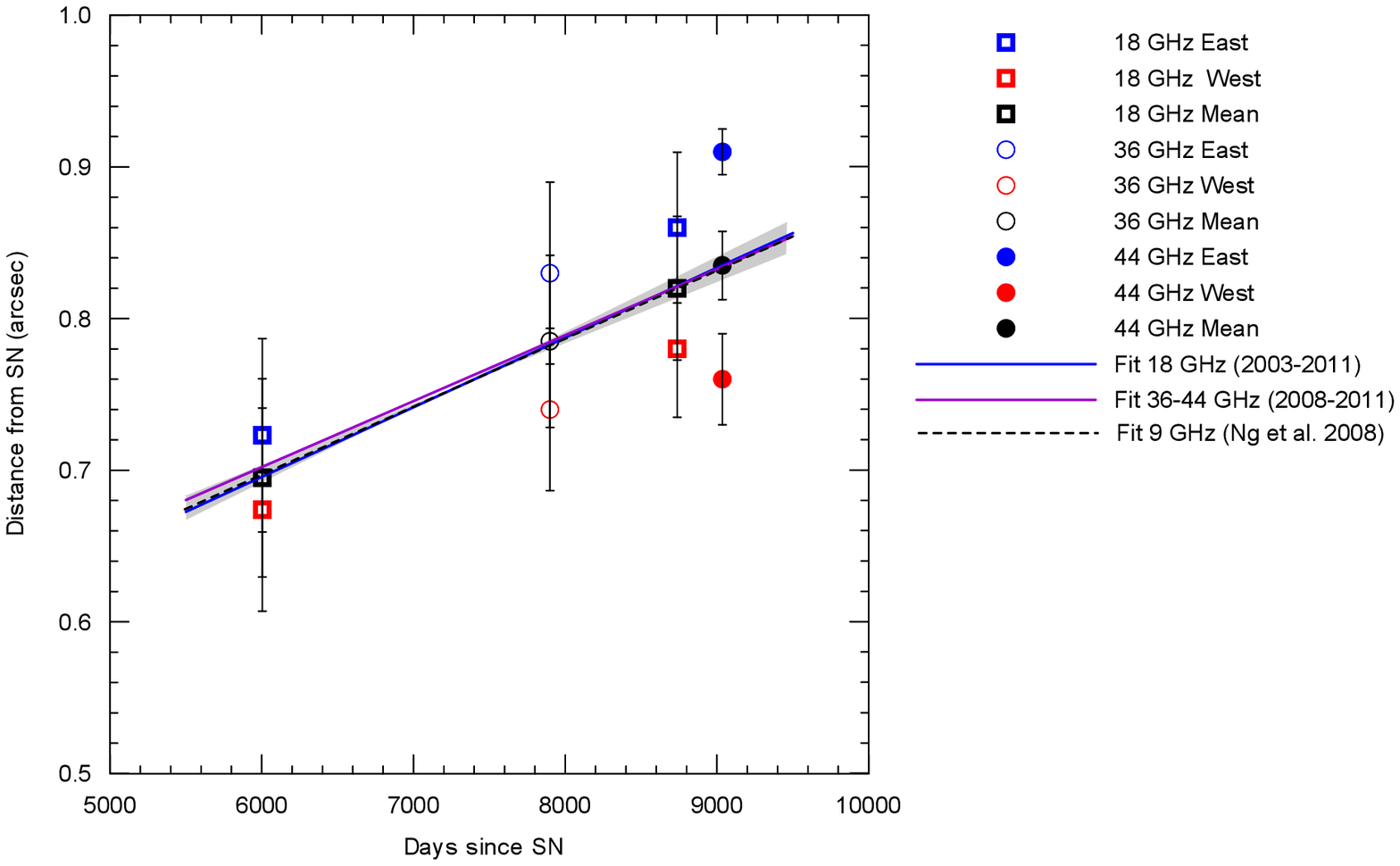}
\caption{Remnant expansion  from images at 18 GHz on day 6003 \citep {man05} and day  8738, at 36 GHz  on day 7815 \citep{pot09} and 44 GHz on day 8886 (this paper). The expansion is estimated from the distance between the VLBI position of the SN \citep{rey95} and the position of the peak of brightness for the eastern and western lobes, as derived from  the PA 90$^{\circ}$ emission profiles centred on the SN coordinates. The blue symbols are associated with the brightness peaks on the eastern lobes, while the red ones represent the distance between the SN position and the brightness maximum in the western lobe.  Mean values derived for each pair of measurements are represented by the black symbols. The  blue line is the linear fit of the measurements at 18 GHz from 2003 to 2011, while the purple line only fits the 36 and 44 GHz within 2008 and 2011. For data between 2008 and 2011, the derived expansion velocity is $3900\pm300$ km s$^{-1}$. As a comparison with the expansion rate derived by \citet{ng08},  the dashed black line is the slope of the linear fit of radius measurements extracted from the 1992--2008 super-resolved images at 9 GHz fitted with a torus model, the error associated with the fit is represented by the grey-filled  area.}
\label{fig:expansion}
\end{center}
\end{figure*}

In Figure~\ref{fig:expansion}, the expansion of the remnant at  18 and 44 GHz is estimated as the change over time of the distance between the VLBI position of the SN \citep{rey95} and the mean position of the peak of brightness for the eastern and western lobes, as derived from  the emission profile at PA 90$^{\circ}$ centred on the SN coordinates.
Comparison between observations at 18 GHz at day 6003 and 8738, gives a velocity of 4100 km s$^{-1}$ (blue line), while the comparison between the 36 and 44 GHz datasets,  within the 2008--2011 time frame,  leads to a velocity of 3900 km s$^{-1}$ (purple line). An identical result of $3900\pm300$ km s$^{-1}$ is obtained if the 2011 dataset at 18 GHz is included in the fit of data between 2008 and 2011, since the 18 GHz measurements align with those at 36 and 44 GHz. All fits are consistent with the 9 GHz results.  

With reference to the SN position, it should be noted that both the datasets at 18 GHz (2003--2011), and the datasets at 36 GHz (2008) and 44 GHz (2011), show an asymmetric expansion of the remnant, with a larger increase of the distances between the SN position and that of the brightness peak on the eastern lobe, compared to the western side. The asymmetry in the expansion has been noted since the images at 9 GHz in early epochs \citep{gae07},  where an asymmetry in the initial expansion of the SN ejecta has been proposed as the explanation for the fact that the eastern lobe is located further than the western one from the SN site. In the case of the 18 GHz datasets, an expansion velocity of 5900 km s$^{-1}$ is derived for the eastern lobe, while  2300 km s$^{-1}$ is the expansion velocity of the western brightness peak. Similarly, in the datasets at 36 and 44 GHz, the eastern expansion velocity is  6300 km s$^{-1}$ while the western is 1600 km s$^{-1}$. Averaging the values, the expansion velocity of the remnant western lobe is 1900$\pm$400 km s$^{-1}$, while that of the eastern lobe is 6100$\pm$200 km s$^{-1}$. While these values are affected by some uncertainty in identifying the location of the remnant with respect to the VLBI coordinates of the SN site, the results from the 18, 36 and 44 GHz data imply that  the expansion velocity of the eastern lobe is significantly  larger than that of the western lobe. Therefore, a marked asymmetry of the remnant emerges not only in terms  of brightness distribution of the radio emission but also in terms of remnant expansion. 

\section{Spectral index measurements}
\label{spectral}

Spectral index maps allow the examination of the spatial variations of the spectral index, $\alpha$ (where $S\propto\nu^{\alpha}$), across the remnant and, therefore, of the electron acceleration processes associated with the propagation of the blast wave.
The 2011 observations at 18 and 44 GHz have been used to derive a map of the distribution of the spectral index in the SNR.  To match the $u-v$ coverage of the 18 GHz observations, the visibilities from baselines greater than 400 k$\lambda$ were filtered out in the 44 GHz datasets. At both frequencies, the images have been derived using an identical reduction procedure. In particular, identical weighting parameters, deconvolution algorithm and phase self-calibration iterations were applied to both datasets.  It should be noted that the self-calibration technique, while improving the image resolution, removes all absolute positional information. Therefore, the astrometry of each image was compared with that prior to self-calibration and the images were shifted to align with recent VLBI observations of the remnant (Zanardo et al., in preparation). This process allowed us to accurately place the position of the remnant with respect to the VLBI position of the SN mentioned in \S~\ref{morphology}. The aligned images were then restored with a $0\farcs4$ circular beam and regridded at a pixel scale of 3 mas. 

\subsection{Methods for measuring the spectral index}
\label{spectral_methods}

To verify that any claimed spectral index variations have not been artificially generated by the data reduction and/or are tied to a particular  spectral mapping technique, the spectral index distribution within the remnant is reconstructed via three different methods: ({\it i}) spectral tomography ($\alpha_{t}$); ({\it ii}) flux ratio ($\alpha_{S}$); ({\it iii}) temperature-temperature (or T-T) plots ($\alpha_{TT}$).

With the method of spectral tomography \citep{kat97}, a difference image, $I_{t}$, is calculated by scaling the 44 GHz image by a trial spectral index, $\alpha_{t}$, and subtracting it from the 18 GHz image, as it follows:
$$
I_{t}(\alpha_{t})\equiv I_{18}-{({\frac{\nu_{18}}{\nu_{44}})}^{\alpha_{t}}}I_{44}
$$
where $I_{18}$ and $I_{44}$ are the images at frequency $\nu_{18}$ and $\nu_{44}$, respectively. When $\alpha_{t}$ reaches the actual spectral index of a particular feature, the feature vanishes in the local background of the difference image. On the other hand, if a component has spectral index greater or smaller than $\alpha_{t}$, the difference image appears with a distinctively positive or negative residual compared to the local background,  as this component will get  over or under-subtracted. 

The second method used to construct a spectral index map is that of direct image division, being  $\alpha_{S}=\log(S_{18}/S_{44})/\log(18/44)$. As this method is sensitive to variations in background/foreground emission that might cause spatial changes in the observed spectral index, it is not usually the preferred method to investigate the spectral index distribution over large SNRs \citep{and93}. 

To validate the results from the flux density ratio $S_{18}/S_{44 }$,  the spectral variations over small regions within the remnant  are also assessed from the flux slope $m$, where $S_{18}=mS_{44}+q$.
This approach was first applied by \citet{cos60} and \citet{tur62} to data of very low angular resolution, where instead of fluxes, the brightness temperatures, $T_{b}\propto\nu^{-(\alpha+2)}$, are plotted at two frequencies, and the spectral index is derived from the best fitting line of the plotted points (thus referred to as temperature-temperature or T-T plot). 

The slope of the best-fitting line to the $T_{b}$ points derived for each pixel of the images at 18 and 44 GHz ($T_{18}$ and $T_{44}$), within a specific region, yields the spectral index $\alpha_{TT}$ of that region. While small differences in the background levels cause spreading of the plotted T-T points, because of the intrinsic spatial averaging of the linear regression method, the regional $\alpha_{TT}$ values are less affected than $\alpha_{S}$ by zero-level data  (\citealp{lea91, zha97}).

\subsection{Interpretation of spectral index variations}
\label{spectral_interpretation}

Understanding the spatial variation of the spectral index and emission across young SNRs is important to investigate the structure of the expanding shock.  The spectral index can be used to probe the compression ratio of the shock on a local scale. In turn, the emission probes cosmic ray and magnetic field density. Combined, these provide the basis for investigating the nature of  shock acceleration processes.

\begin{figure*}[htp]
\begin{center}
\vspace{0.0mm}
\advance\leftskip-3.5mm
\includegraphics[width=150mm,angle=0]{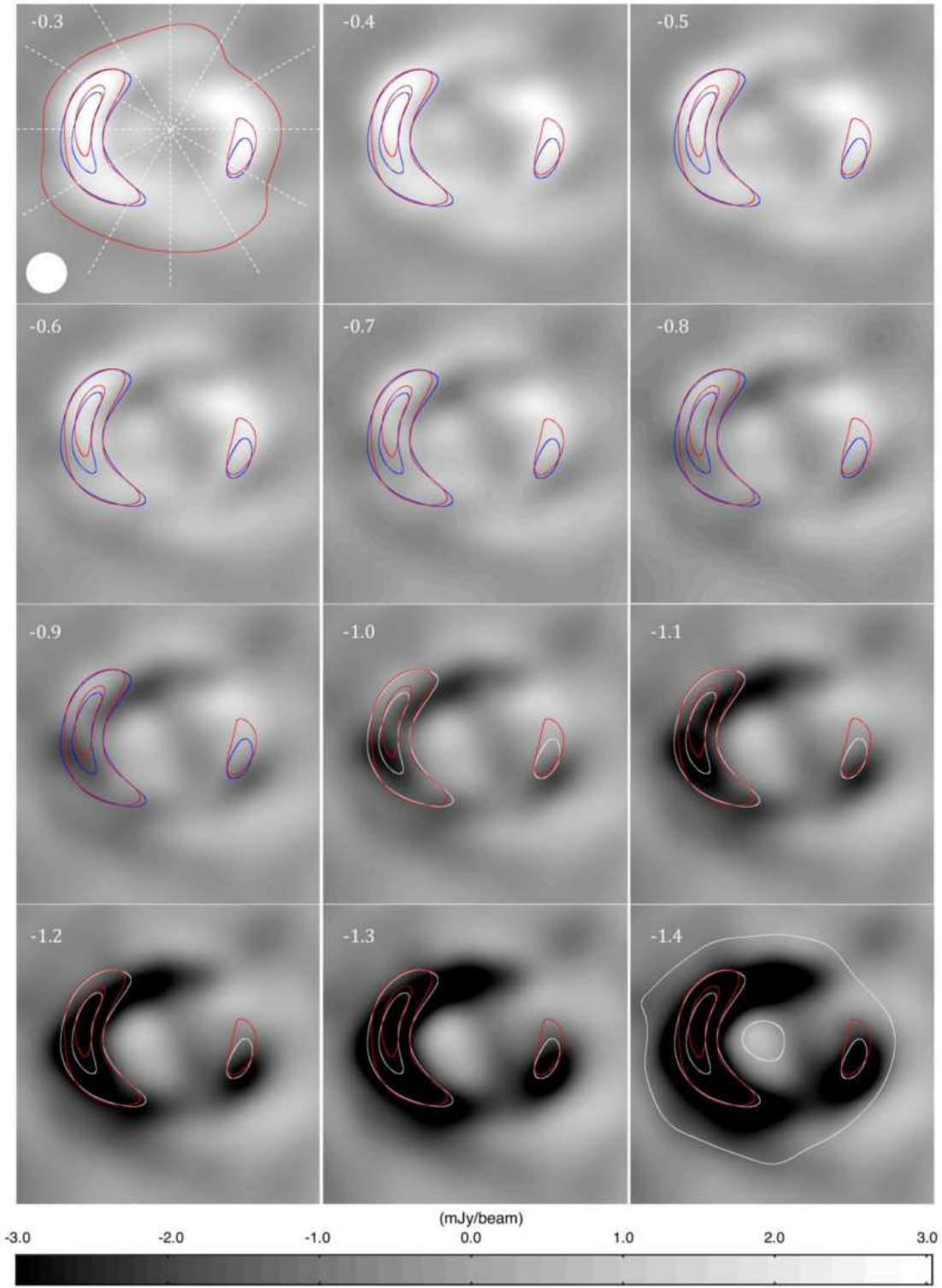}
\caption{Tomographic spectral images of the remnant of SN 1987A, where the spectral index is defined as $S\propto\nu^{\alpha}$. The spectral images are derived from the 18 GHz image from observations performed on 2011 January 26, and the 44 GHz image from observations between in 2011 January and November. Both images are centred on the VLBI position determined by Reynolds et al. (1995), and restored with a circular $0\farcs4$ beam (top panel, lower left corner).  The grey scale shows difference images for $-1.4\le\alpha_{t}\le-0.3$, in steps of 0.1. Blue/white contours correspond to the 44 GHz data convolved with a $0\farcs4$ circular beam at levels 60\%--90\% of the peak flux density, while  the related contours corresponding to the 18 GHz image are in red. In the top left panel, axes at six position angles, centred on the SN position, are also indicated. All images are set on the same intensity range.}
\label{fig:spectral_18-44_tomography}
\end{center}
\end{figure*}

In the tomography gallery in Figure~\ref{fig:spectral_18-44_tomography}, spectral index variations are investigated in the range between $-1.4\le\alpha_{t}\le-0.3$, with increment $\delta\alpha_{t}=0.1$. It can be seen that the spectral index distribution across the remnant is not uniform. 
In detail, the inner areas of the remnant located along PA $\sim$30$^{\circ}$ (210$^{\circ}$), appear to match the grey background for $-0.6\le\alpha_{t}\le-0.3$, with $\alpha_{t}\sim-0.4$ in the inner part of northern region, and $\alpha_{t}\sim-0.5$ in the inner part of the feature located slightly SW of the SN position. The spectral index gradually steepens towards the edges, with the northern and southern regions becoming over-subtracted at  $\alpha_{t}\sim-0.7$ and $\alpha_{t}\sim-1.0$, respectively.
For the bright regions on the eastern lobe of the remnant, sites within the contours at 90\% of the peak flux density (see  Figure~\ref{fig:spectral_18-44_tomography}) primarily match the grey background for $\alpha_{t}\sim-0.8$. Within the 60\% contours, $\alpha_{t}$ changes from $-0.7$ (southern part), to $-0.8$ (central region) and $-0.9$ (northern region). The entire bright area on the eastern lobe is over-subtracted for $\alpha_{t}\sim-1.1$. 
On the western lobe, sites within the 60\% contours have a wider range than on the eastern lobe, as $\alpha_{t}$ varies from $-1.1$ to $-0.7$ with a north-south gradient.  
Southern regions in the tomography maps disappear into the background for $-0.8\le\alpha_{t}\le-0.7$, with larger errors on sites close to the edges of the 10\% flux density contour of the 44 GHz image. Larger spectral index variations can be seen in the N-NW sites, with $-0.9\le\alpha_{t}\le-0.6$.
The sites that can be associated with spectral index $-1.4\le\alpha_{t}\le-1.2$ correspond to flux density at 44 GHz close to zero and, therefore, are more affected by error in the image subtraction.

The spectral index map, as derived from the direct division of the images at 44 and 18 GHz, is shown in Figure~\ref{fig:spectral_18-44_divide}. 
It can be seen that the spectral indices primarily vary in the range $-0.9\le\alpha_{S}\le-0.7$. 
In particular, on the eastern lobe, within  the 90\% contours of the peak flux density at both 18 and 44 GHz, $-0.9\le\alpha_{S}\le-0.7$, while slightly steeper values, $-1.0\le\alpha_{S}\le-0.8$, correspond with the western brightest sites. 
On both lobes, near the northern edge of the 60\% contours, the spectral indices peak at $\sim-0.9$ on the eastern sites and $\sim-1.1$ on the western ones. Flatter spectral indices ($-0.6\le\alpha_{S}\le-0.3$) can be located within the central and northern regions of the remnant at $\sim$PA 30$^{\circ}$.

It is noted that the regions at the edges of the $\alpha_{S}$ map, i.e.\ $\alpha_{S}\lesssim-1.2$ (in dark-blue and black) and $\alpha_{S}\gtrsim-0.3$ (in light-blue and white) are likely affected by higher uncertainty, since these values are associated with lower S/N ratios in either the 44 or 18 GHz images.
The effects of low S/N  on the $\alpha_{S}$ distribution can be assessed from the histogram of the spectral indices over the entire map. In Figure~\ref{fig:Histo_tot}, the black curve is the Gaussian fit over the entire $\alpha_{S}$ range, which yields median value $\alpha_{S_\mu} =-0.78$ and standard deviation $\sigma=0.39$. Spectral indices $\alpha_{S}\le\alpha_{S_\mu}-\sigma$ correspond to regions in the 44 GHz image  of low flux density (S/N$<$100), while $\alpha_{S}\ge\alpha_{S_\mu}+\sigma$ are associated with regions of the 18 GHz image where S/N$<$100.  In  Figure~\ref{fig:Histo_tot}, the blue curve is a better-fitting Gaussian function for $\alpha_{S_\mu}-\sigma\le\alpha_{S}\le\alpha_{S_\mu}+\sigma$. 

%
%
\begin{center}
\begin{deluxetable}{ccccc}
\tablecaption{Regional Spectral indices  \label{tab_alpha}} 
\tablewidth{3.13in}
\tablecolumns{5}
\tablehead{
\colhead{Region\tablenotemark{$a$}} 										&  
\colhead{$\alpha_{TT}$\tablenotemark{$b$}} 									& 	
\colhead{$\alpha_{S_\mu}$\tablenotemark{$c$}} 								&
\colhead{$\Delta\alpha$\tablenotemark{$d$}} 									&
\colhead{$\Delta\alpha/\alpha_{S_\mu}$\vspace{1.0mm}}\\
\colhead{} 	&  \colhead{} 			& 	\colhead{} 			& 	\colhead{}  & 		\colhead{(\%)}
}
\startdata
N1			&	$-0.88\pm0.13$	&	$-0.67\pm0.05$ 		&	$0.21$ 	   &		$31$\\
N2			&	$-0.65\pm0.25$	&	$-0.71\pm0.05$		&	$0.06$ 	   &		$8$\\	
E1			&	$-0.91\pm0.07$	&	$-0.79\pm0.06$		&	$0.12$ 	   &		$16$\\	
E2			&	$-0.80\pm0.06$	&	$-0.66\pm0.05$		&	$0.14$ 	   &		$22$\\	
S1			&	$-0.85\pm0.09$	&	$-0.73\pm0.05$		&	$0.12$ 	   &		$16$\\	 
S2			&	$-0.72\pm0.09$	&	$-0.85\pm0.06$		&	$0.13$ 	   &		$15$\\	
W1			&	$-0.76\pm0.16$	&	$-1.09\pm0.08$		&	$0.33$ 	   &		$31$\\	 
W2			&      $-0.80\pm0.11$		&	$-0.75\pm0.05$		&	$0.05$ 	   &		$6$\\	 
C1			&	$-0.52\pm0.18$	&	$-0.80\pm0.06$		&	$0.28$ 	   &		$35$	 
\enddata
\tablenotetext{$a$}{The regions selected for the T-T plots are designated in Figure~\ref{fig:T-Tmap}.}
\tablenotetext{$b$}{The errors on $\alpha_{TT}$ are the combination of the $1-\sigma$ error on the slope of the linear fit and the uncertainty in the flux calibration.}
\tablenotetext{$c$}{The median spectral index, $\alpha_{S_\mu}$, is derived by fitting a Gaussian to the histogram of the $\alpha_{S}$ values in the region (see Figure~\ref{fig:T-Tplots}). The errors are derived from the flux calibration uncertainty.}
\tablenotetext{$d$}{$\Delta\alpha=\lvert\alpha_{TT}-\alpha_{S_\mu}\rvert$. Note that the errors on $\alpha_{TT}$ and $\alpha_{S_\mu}$ are correlated.}
\end{deluxetable}
\end{center}

\vspace{-4.5mm}
The T-T plots are used to measure the dominant spectral index within $0\farcs5\times0\farcs5$ squared regions, as designated in Figure~\ref{fig:T-Tmap}. The resultant $\alpha_{TT}$ values are listed in Table~\ref{tab_alpha}, while in Figure~\ref{fig:T-Tplots} the T-T plots for each region are shown together with the histograms of the corresponding $\alpha_{S}$ distribution. In seven of the selected areas $-0.91\le\alpha_{TT}\le-0.72$,  whereas flatter spectral indices are derived for the central region C1 ($\alpha_{TT}\sim-0.52$), which includes the VLBI position of the SN, and the adjacent northern region N2 ($\alpha_{TT}\sim-0.65$).  The larger differences in the temperature levels within C1 and N2 (see Figures~\ref{fig:T-Tplots}), can be seen in the spread of the T-T points, which results in the larger error bars attached to the linear fit.  As shown in in Figure~\ref{fig:T-T_tot}, the spectral index derived for the entire remnant is $\alpha_{TT_m}=-0.80\pm0.05$, where the flux calibration uncertainty is factored in the error (see \S~\ref{New_obs}).

As mentioned in \S~\ref{spectral_methods}, among the three methods used to assess the spectral index, the T-T plot is  the more robust, since the spectral index  variations resulting from the image division and subtraction depend, to different extent, on the local intensity of the flux density at the two frequencies. However, since the T-T method requires boxes larger than the angular resolution of the images, the characteristic spectral index of smaller regions, such as the features of flatter $\alpha_{t,S}$ at PA $\sim30^{\circ}$, is likely influenced by surrounding steeper emission. Otherwise, when larger structures are analysed, such as the sites of brighter emission on the eastern and western lobes, the spectral indices derived from the T-T plots are likely more accurate than a local average of $\alpha_{t}$ and $\alpha_{S}$. As listed in Table~\ref{tab_alpha}, over the brightest sites of the remnant the discrepancy between $\alpha_{TT}$ and regional $\alpha_{S_\mu}$, which is the median value of the $\alpha_{S}$ distribution within the T-T region, is around 19\% over the eastern lobe and peaks to $\sim$31\% in W1. 

The average spectral index of the western bright regions (W1--W2) at $\sim-0.78$ matches $\alpha_{TT_m}$. As discussed earlier, the steeper $\alpha_{S}$ values in W1 are likely affected by the higher noise levels in the adjacent northern region. Over the eastern lobe, the T-T plots for regions E1 and E2 give an average spectral index of $-0.86$, which is steeper than $\alpha_{TT_m}$. The steepest value  $\alpha_{TT}\sim-0.91$ is associated with region E1, which covers the brightest sites of the remnant.

From the theory of diffusive shock acceleration (DSA) of energetic particles in a uniform magnetic field \citep[see][and references therein]{dru83}, the standard compression ratio, $r$, can be derived from the radio synchrotron spectral index. The $\alpha_{TT}$ values given in Table~\ref{tab_alpha} yield similar compression ratios  in regions W1, W2  and E2, specifically with $r_{W1}=2.97\pm0.43$,  $r_{W2}=2.87\pm0.26$ and $r_{E2}=2.87\pm0.14$, while a lower compression ratio, $r_{E1}=2.65\pm0.13$, results in region E1.  
These results suggest a particle spectrum harder over the western regions than on the brightest eastern sites. For a particle density expressed as $n(p)\propto p^{-\gamma}$, where {\it p} is the scalar  momentum, in the scenario of ordinary diffusion, $n_{(W1-W2)}\propto p^{-2.56\pm0.26}$ is the spectrum of the western side of the SNR, while $n_{(E1)}\propto p^{-2.82\pm0.14}$ is obtained for region E1.

The lower compression ratio in region E1 could correspond with the asymmetric expansion observed in the remnant, which appears to be driven by higher eastern velocities (see \S~\ref{expansion}). As $r = \rho_{2}/\rho_{1}$ \citep{jon91}, where  $\rho_{2}$ and $\rho_{1}$ are the downstream and upstream gas densities, respectively, a lower compression ratio over the eastern brightest sites could be due to the higher-velocity shock drifting the cosmic rays (CR) upstream.
The CR diffusion could induce gas heating upstream of the shock \citep{ptu10} and, thus, a higher upstream pressure. 
If the SNR magnetic field is tangled rather than uniform \citep{duf95}, the softer spectrum could be the result of local sub-diffusive particle transport, as the accelerated particles get partially trapped in the vicinity of the shock front by structures in the magnetic field  \citep{kir96}. In this case, the particle spectrum corresponding to region E1 would be steeper than for diffusive transport, with $n_{(E1)}\propto p^{-3.35\pm0.19}$, where $\gamma=\gamma_{DSA}(1+0.5/r)$ \citep{kir96}.

The softer spectrum in E1 likely implies that a lower fraction of injected particles, i.e.\ a higher injection efficiency, is required for the high emission observed in the eastern region. 
If the stronger eastbound shock is coupled with a more efficient particle acceleration, the magnetic-field on the eastern lobe might become locally amplified (\citealp{bel01,bel04}). In fact, according to \citet{bel01}, with  efficient CR acceleration, $(\Delta B/B)^{2} \approx \mathcal{M}$, where $B$  and $\Delta B$ are the background  
and fluctuating magnetic fields, respectively, and $\mathcal{M}$ is the Mach number of the shock. Thus, a higher $\mathcal{M}$ would likely induce non-linear amplifications of the magnetic field generated in the SNR. As the magnetic-field amplification depends on the magnetic field orientation, it can be noted that in asymmetric bipolar (or bilateral) remnants, such as SNR 1987A, regions of higher emission  have been linked to either  quasi-parallel or quasi-perpendicular magnetic field inclinations  (\citealp{ful90, gae98, orl07}). 

As regards the $\alpha_{TT}$ results in the northern and southern parts of the SNR, adjacent small-scale features of both steeper and flatter spectra, lead to spectral indices flatter than $\alpha_{TT_m}$ in the S2 and N2 regions, while the values derived for regions N1 and S1 seem still dominated by the steeper indices associated with the bright sites on the eastern lobe. We note that, overall, the spectral indices over the high-emissivity regions are steeper than predicted by first-order Fermi acceleration at a strong shock. This agrees with studies on the spectral distribution of young and rapidly-evolving remnants \citep[e.g.\ see discussion on Cas A in][]{and91}, where, since concurrent high-emissivity and steep spectral indices cannot be obtained from first-order acceleration models, other mechanisms, such as CR-mediated shocks and turbulent acceleration, are invoked \citep{and93}.

In both $\alpha_{t}$ and $\alpha_{S}$ distributions, gradual transitions from steeper to flatter spectral indices (i.\ e. $-0.8\lesssim\alpha_{t,S}\lesssim-0.5$) can be seen on the edges of the remnant and around, or in proximity, of the brightest sites.
Spectral index gradients in older and larger SNRs have been explained as due to multiple overlapping spectral index structures (\citealp{del02, tam02}).  According to hydrodynamic simulations coupled to DSA by \citet{ell07} of a spherical SNR, a spectral gradient might reflect a  spectral index structure over multiple concentric shells of shocked material within the forward and the reverse shocks, with a CR population varying  between the concentric shells, such that the spectral index results flatter on the outer shell than on the inner ones. The scenario of sub-diffusive  transport of particles at the shock front, also leads to small-scale spectral gradients in the downstream plasma, which would correspond to 
the transition between sub-diffusive and diffusive behaviour of the particle propagation \citep{kir96}. 

All methods used for spectral analysis show two inner regions of flatter spectral indices, where  $-0.6\lesssim\alpha_{TT}\lesssim-0.5$ and $-0.5\lesssim\alpha_{t,S}\lesssim-0.3$. The more central feature, which overlaps with the SN site (see Figure~\ref{fig:spectral_18-44_divide}), has $\alpha_{t}\approx-0.5$, $-0.6\lesssim\alpha_{S}\le-0.5$ and $\alpha_{TT}\sim-0.52\pm0.18$. The second feature, located further north, can be associated with  $\alpha_{t}\approx-0.4$ and $-0.4\lesssim\alpha_{S}\lesssim-0.3$, while the corresponding $\alpha_{TT}$, between regions N1 and N2, is affected by the nearby steeper spectra.

We note that a central feature of flatter spectrum has been seen in other two-frequency spectral index images of the SNR (\citealp{pot09,lak12}).
In Figure~\ref{fig:spectral_18-44_divide} the spectral index image is also overlaid with the contours of the  {\it HST} observations in 2011\footnote[2]{STScI-2011-21, NASA, ESA, \& Challis P. (Harvard--Smithsonian Center for Astrophysics)
\begin{tabular}{l p{88mm}}
\url{http://hubblesite.org/newscenter/archive/releases/2011/21/image/}
\end{tabular}} (in black),
and with the 10\% level of the peak flux density of the high-resolution image at 44 GHz (in yellow). It can be seen that the central feature of flatter electron spectrum, not only coincides with the inner region of fainter emission that is visible in the radio at higher resolution, but also  partially overlaps with the western side of the optical ejecta.
On the contrary, the northern feature, which appears to connect with the central feature along PA $\sim$30$^{\circ}$, approximately in alignment with the main direction of the optical ejecta, does not seem to have been noted before. This feature exhibits $\alpha_{t,S}$ values flatter than the central region, and, since it  appears to be smaller than the $0\farcs5\times0\farcs5$ boxes used for the T-T plots,  cannot be associated with a local $\alpha_{TT}$ value.
Nevertheless,  since it corresponds to a site characterised by high flux density both at 18 and 44 GHz, the related $\alpha_{t}$ and $\alpha_{S}$ values are unlikely affected by significant error.

While pulsar wind nebulae (PWNe) in the radio band are identified by $-0.3\lesssim\alpha\lesssim0$ \citep{gae06}, the acceleration origin in the above discussed features of $-0.5\lesssim\alpha\lesssim-0.3$, could be explained in terms of injection by a pulsar. 
Further observations at high frequencies and monitoring of any morphology changes is required to clarify the nature of these features. 
\vspace{0.0mm}

\section{Emission at other wavelengths}
\label{other_wavelenghts}

As shown in early models of Type II SNRs \citep{che82}, the SN-CSM interaction in SNR 1987A has generated a main double-shock structure, which includes, from the outside inwards, the forward shock and the reverse shock. Between the forward and reverse shocks, reflected shocks should be present due to the forward blast wave encountering the high-density CSM in the ER \citep{bor97}. While the structure of the shock can be probed by comparing the remnant morphology in the radio and X-ray wavelengths, the distribution of the progenitor dense material can be investigated via H$\alpha$ images.

The radio emission mainly originates at the forward shock as this collides with the dense CSM associated with the ER, thus creating a  discontinuity in the magnetic field where particles  are accelerated \citep{zan10}. The forward shock can be identified with the sharp outer edge of the SNR shell,  which  propagates into the CSM of spatially varying density. 
A similar scenario happens at the reverse shock, or inner edge of the SNR envelope, which is normally dominated by the X-ray emission.  Between the reverse shock and the ER, the reflected shocks  lead to the higher temperatures of the shocked gas and a flatter density profile. 
Up to day $\sim$4000, the X-ray emission has likely been generated by the interactions with the low-density H{\sc ii} region located on the inside of the ER \citep{che95},  whereas, in the last $\sim$5000 days, it has primarily  originated from the interaction  with the dense inner ring \citep{par11}. 

In Figure~\ref{fig:44_X-ray_polar_PA}, the 44 GHz image is compared to recent X-ray observations \citep{hel12}. As noted by \citet{ng09}, the X-ray emission exhibits an east-west asymmetry less marked than what seen in the radio, while the north-south asymmetry is significant. 
In particular, the brightest X-ray site, which  appears in the NE quadrant between PA 30$^{\circ}$ and 60$^{\circ}$, overlaps with the brighter regions on the eastern lobe of the 44 GHz image, but peaks  slightly northwards of the eastern radio peak.  In the western lobe, the X-ray emission peaks at PA $\sim$270$^{\circ}$ and between PA 300$^{\circ}$ and 330$^{\circ}$. While the radio image is also reaching the western peak of brightness at PA $\sim$270$^{\circ}$, the region between PA 300$^{\circ}$ and 330$^{\circ}$ corresponds to one of the fainter sites (see \S~\ref{morphology_44}). 
The offset between the X-ray and radio peaks is demonstrated in the polar projection of the two images shown in Figure~\ref{fig:44_X-ray_polar_PA}. 
 
The 44 GHz map and the {\it HST} image from recent observations are superimposed in Figure~\ref{fig:44+Hbl}. It can be seen that the morphology of the radio emission matches the ring optical features. As shown in Figure~\ref{fig:spectral_18-44_divide}, on both the east and west sides of the optical ring, the easternmost and westernmost  hot spots appear to overlap with the bright regions on the eastern and western lobes of the radio images. 
The fainter regions of the ring visible at 44 GHz, i.\ e. located in the NW and SW quadrants, seem to coincide with sites where clusters of  H$\alpha$ hot spots appear somewhat disjointed. This is especially pronounced on the NW quadrant, between PA 300$^{\circ}$ and 330$^{\circ}$, where the forward shock might have overtaken the ER and reached less dense CSM regions, while the inner surface of the ring is still hot enough to emit X-rays.

As regards the remnant asymmetry, contrary to the radio images, the optical emission now appears markedly brighter on the western lobe. The difference in the asymmetry direction between the nonthermal radiation and the H$\alpha$ emission is  more clear in the RGB overlay shown in Figure~\ref{fig:RGB:44+Hbl+X-ray2}. It can be noted that the eastern lobe of the ring is dominated by the radio ({\it red}) and X-ray ({\it blue}) emissions, which turn into shades of violet in the overlay, especially in the NE quadrant where the radio and X-ray brighter sites overlap. On the western lobe, the fainter radio emission is now overshadowed by the H$\alpha$ emission ({\it green}).  
This appears to validate the hypothesis that the remnant asymmetry is likely due to an asymmetric explosion of the progenitor rather than to an asymmetric distribution of the CSM.  
We may therefore expect the radio and X-ray emission to follow suit, with the east-west asymmetry gradually reversing.
The possibility of an asymmetric initial explosion was suggested by \citet{che89} to explain the asymmetric expansion of the remnant, whereas hydrodynamic calculations of the remnant evolution from the asymmetric explosion of a $\sim$20 M$_{\sun}$ merger of binary systems, have been proposed to explain the BSG progenitor and to link the BSG-RSG wind interaction to the triple-ring nebula structure (\citealp{pod07, mor07, mor09}).

In Figure ~\ref{fig:44+Hbl}, the radio emission from the inner region of the remnant as seen at 44 GHz can be compared to the optical emission from the ejecta. It is noted that the ejecta is  blue and red-shifted along the line-of-sight \citep{kja10} and, like the emission from the ER, the optical ejecta is characterised by an east-west asymmetric morphology,  whose origin is yet unclear \citep{lar11}.
The western side of the ejecta appears to overlap with the eastern part of the central feature visible at 44 GHz and, as mentioned in   \S~\ref{spectral_interpretation}, extends over a region of the SNR that can be associated with flatter spectral indices.  Moreover, the `hole' in the ejecta, which is close to the SN location (see Figure~\ref{fig:spectral_18-44_divide}) and  has become more pronounced  in recent years  \citep{lar11}, also overlaps with the central emission detected at 44 GHz.
\vspace{-2.0mm}

\section{Free-free absorbed compact source}
\label{free-free}

\citet{mat11} have estimated that the largest dust component, with mass  $0.4-0.7\, M_{\sun}$, sits in the central region of the SNR.  However \citet{lak12} have concluded that, for observations up to 94 GHz, the dust does not contribute to the radio emission. The central feature detected at 44 GHz might therefore originate from a compact source or PWN located in the vicinity of the SN ejecta. 

To determine whether non-thermal radio emission from a pulsar or a PWN located in the inner regions of the SNR would be detectable, we estimate the fraction of this emission that would be absorbed by the ionised gas likely located within the ER.
Assuming the ionised mass in the ejecta, $M_{ej}$, to be distributed within a sphere of radius $R$, the free-free optical depth, $\tau_{ff}$, is proportional to the path length through the free-free absorbing spherical region along the line-of-sight (los). In convenient astronomical units, $\tau_{ff}$ can be estimated as  \citep{ryb79}

\vspace{-4.0mm}
 $$\tau_{ff}\approx\tau_0\,T^{-3/2} \nu^{-2} \,\overline{g_{ff}} \,EM$$
where 
$T$ is the temperature of the plasma in K, $\nu$ is the frequency of the flux density in GHz, $\overline{g_{ff}}=\overline{g_{ff}}(\nu,T)$ is the velocity-averaged Gaunt factor appropriate to the observing frequency, and  the emission measure is defined as
$$EM\equiv \int_{\rm los}  \! n_{e}n_{i} \, \mathrm{d} l$$ 
where  $n_{e}$ and $n_{i}$ are the electron and ion densities, respectively, expressed in cm$^{-3}$,  and $l$ is the path length,   expressed in pc, along the line-of-sight.
As H{\sc ii} regions typically have electron temperatures of order 10$^{4}$ K, indicated by $T_{4}$, and, in the radio regime,  
$\overline{g_{ff}}(\nu,T)\approx 5.99 \,T_{4}\,^{0.15} \nu^{-0.1}$ \citep{bro87}, 
 the free-free opacity becomes
\vspace{1.0mm}
$$ \tau_{ff}\approx3.28\times10^{-7} \,T_{4}^{-1.35} \nu^{-2.1} n_{e}^{2} R$$
assuming $n_{e}= n_{i}$, which is approximated by $n_{e}(M_{ej},R)\approx 3\,M_{ej}(m_{p}\,4\pi \,R^{3})^{-1}$, and  $\int_{\rm los} \mathrm{d} l \approx R$. \\

\vspace{-6.0mm}
Figure~\ref{fig:ff_absorb} shows the variation of $\tau_{ff}$ as a function of the ionised mass for $M_{ej}\lesssim10\,M_{\sun}$ and  volume size  
$R\le0. 1$ pc. The contour levels are plotted for $0\le\tau_{ff}\le2$. 
According to this simplified model,  an ionised mass in excess of $5\,M_{\sun}$ within a radius of 0\farcs2 (approximately corresponding to the size of the densest part of the ejecta visible with {\it HST}), is required to produce a free-free optical depth of unity at 44 GHz. 
We note that the flux density from the emitting plasma, within a $\sim0\farcs2$ solid angle,  given by $S_{M_{ej}} = (2k_{B}T / \lambda^{2}) \Omega$, is negligible.
In this scenario, the radiation observed through the region of ionised plasma would be $S_{obs} \approx0.63\,S_{44c}$, where $S_{44c}$ is the actual flux density at 44 GHz associated with the compact source in the background.
Therefore, since the flux density measured in the inner region of the 44 GHz image is $S_{obs}\sim1.4$ mJy (see \S~\ref{morphology_44}),  the actual radio emission originating from a compact source inside the ER could be as high as $\sim$ 2.2 mJy. Using the spectral index derived in Figure ~\ref{fig:spectral_index}, this emission would lead to $S_{94c}\sim 1.3$ mJy at 94 GHz, which is consistent  with the upper limit of 1 mJy set by \citet{lak12} at that frequency, for any discrete radio source in the central region of the remnant.

In the case of a more realistic,  steeper density profile, such as that  proposed by \citet{che89} with $n_{e}\propto R^{-9.6}$,  the emission from a PWN would be more heavily absorbed and could be detected only if there are holes in the ejecta. As it has been noted from {\it HST} images in the R-band and B-band, the ejecta  is likely to have holes  \citep{lar11} (see  Figures~\ref{fig:44+Hbl} and \ref{fig:RGB:44+Hbl+X-ray2}), thus the radiation emitted by a PWN or compact source might escape the dense ionised material and become visible  at lower frequencies. On the other hand, with a more compact density distribution, part of $M_{ej}$ may have fallen back onto the neutron star that formed in the explosion, and, subsequently, the neutron star and any accreted ejecta might have already collapsed into a black hole.

\section{Conclusions}
\label{conclusions}

We have presented the first image of SNR 1987A at  44 GHz, as derived from ATCA observations performed in 2011.
To investigate the spectral index distribution across the remnant, this new image has been analysed in conjunction with that derived from contemporaneous observations at 18 GHz.  The emission morphology has been also compared to contemporaneous optical and X-ray observations. A summary of our findings is as follows:
\begin{itemize}
\item{Consistent with previous radio observations, the 44 GHz image shows a marked asymmetry in the emission distribution.  More specifically, the east-west asymmetry ratio is $\sim1.5$ from the ratio of the brightness peaks in the radial profiles at PA $\sim$90$^{\circ}$, is $\sim1.6$ from the integrated flux densities over the eastern and western halves of the image.
These values are higher than the $\sim$1.4 ratio derived for the new 18 GHz image and the ratio previously measured with images at lower frequencies.}
\item{The comparison between the new images at both 18 and 44 GHz with corresponding observations performed in earlier epochs, specifically the  2003 observations at 17 and 19 GHz and the 2008 observations at 36 GHz, highlights an asymmetric expansion of the remnant, with expansion velocities on the eastern lobe significantly higher than what measured on the western lobe.}
\item{The 18--44 GHz spectral index distribution is measured at an angular resolution of $0\farcs4$. The spectral indices in SNR 1987A primarily range between $-1.1$ and $-0.3$, with a mean of $-0.8$. Spectral indices associated with the brightest sites over the eastern lobe are steeper than the mean value.  The steeper spectrum on the eastern lobe implies  compression ratios slightly lower than on the western bright sites,  and could be correlated with the higher expansion rate measured on the eastern side of the remnant. Two regions of flatter spectral indices are identified, one approximately located in the centre of the SNR and the other located further north. These two features lie at PA $\sim30^{\circ}$.}
\item{There is a strong correspondence between major features of the emission at 44 GHz, and the arrangement of the hot spots shown in the H$\alpha$ emission. The direction of the east-west asymmetry of the X-ray and radio emission, is opposite to that of the H$\alpha$ emission. This fact supports the hypothesis that the remnant asymmetric morphology might be due to an asymmetric explosion, rather than to an asymmetric distribution of the CSM.}
\item{At 44 GHz, a central feature of fainter emission appears to extend over the SN site, and to overlap with the western side of the ejecta as seen by {\it HST}. This feature corresponds to a region of flatter spectral indices in the 18--44 GHz spectral map, which could indicate the presence of a compact source or a PWN. The origin of this emission is unclear.
However, simple free-free absorption models suggest that the radiation emitted by a compact source inside the equatorial ring may now be detectable at high frequencies, or at lower frequencies if there are holes in the ionised component of the ejecta. 
Future high-resolution observations, both at lower frequencies with VLBI and at higher frequencies with ATCA and the Atacama Large sub-Millimeter Array,  will be crucial to further investigate the nature of this emission.}
\end{itemize}

\vspace{-4.0mm}

\begin{figure*}[!htp]
\begin{minipage}[t]{160mm}
\begin{center}
\vspace{0mm}
\advance\leftskip23.5mm
\includegraphics[width=113.5mm, angle=0]{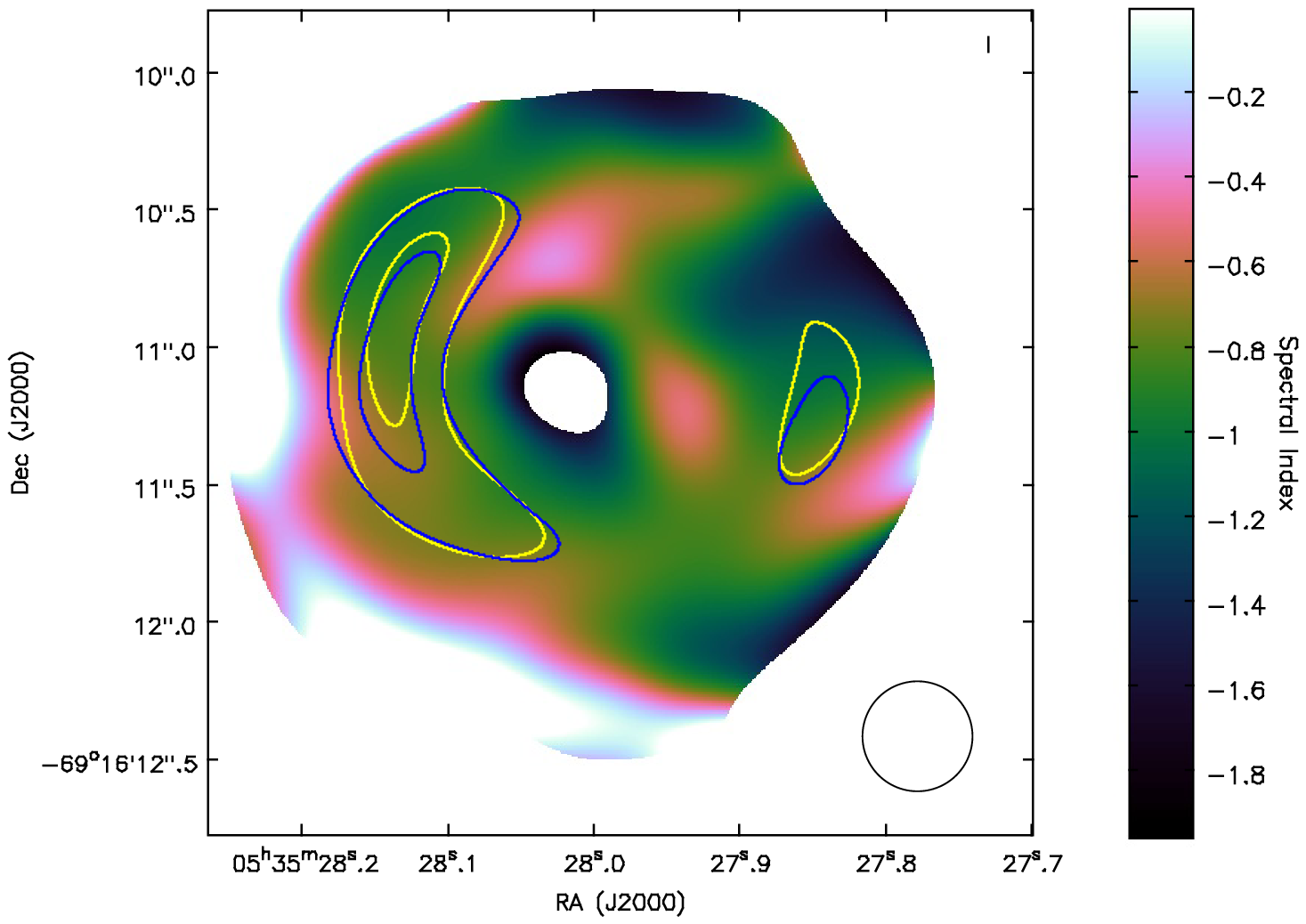}
\end{center}
\end{minipage}
\begin{minipage}[c]{160mm}
\begin{center}
\vspace{3mm}
\advance\leftskip+30.3mm
\includegraphics[width=104.5mm, angle=0]{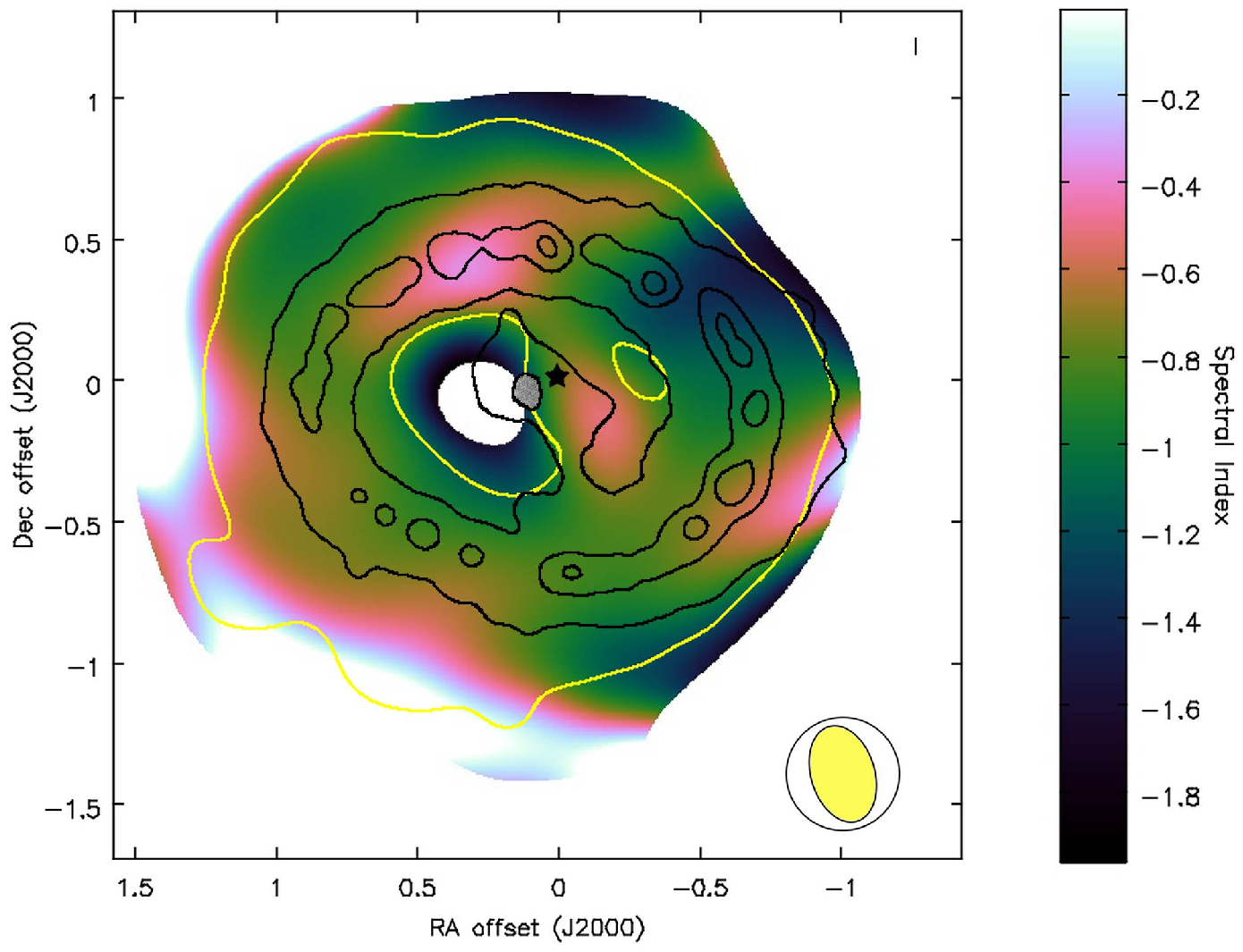}
\end{center}
\end{minipage}
\caption{ 
{\it Top}: Map of the $18-44$ GHz spectral index distribution. Both the 18 and 44 GHz Stokes-I images have been restored with a $0\farcs4$ circular beam,  and regridded at a pixel scale of 3 mas. Image regions below the 44 GHz rms noise level were masked. Contours of the  60\% and 90\% flux density levels for the 18 GHz ({\it yellow}) and 44 GHz ({\it blue})  images, identify the emission peaks  on each lobe. 
{\it Bottom}: To  locate the inner regions of the spectral index map with respect to the emission measured at 44 GHz, the outline of the diffraction-limited image is shown via its contour at 10\% of the peak flux density ({\it yellow}). 
To locate the density distribution of the CSM, the spectral index map is also overlaid with the contours of the {\it HST} image derived from observations performed in 2011\protect\footnotemark[2], specifically at the 2\%, 7\%, 12\% and 20\% emission levels 
({\it black}). As indicated by \citet{lar11}, the innermost {\it HST} contour of the ejecta ({\it gray fill}), east of the SN position ({\it black star}), corresponds to the location of an emission drop, or `hole', in the optical  image.
}
\label{fig:spectral_18-44_divide}
\end{figure*}

\begin{figure}[!tp]
\begin{center}
\vspace{-0.0mm}
\advance\leftskip-4mm
\includegraphics[width=95mm,angle=0]{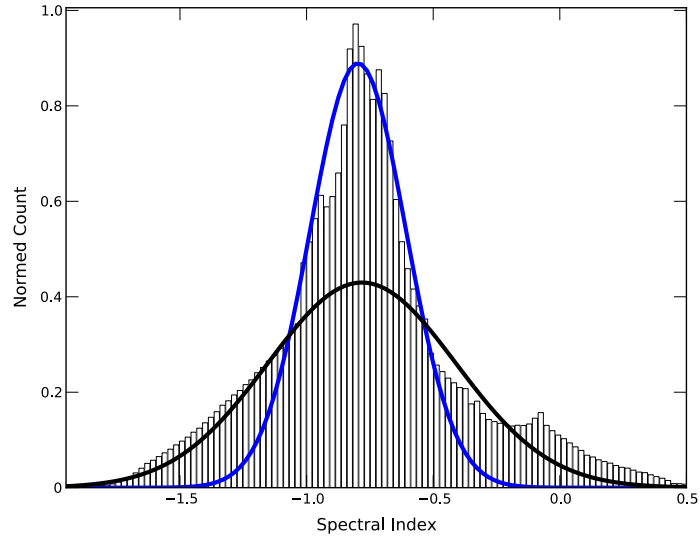}
\caption{Histogram of the spectral index values, $\alpha_{S}$, derived from direct division of the images at 44 and 18 GHz (see Figures~\ref{fig:spectral_18-44_divide}). The vertical axis is the normalised number of occurrences. The black line is the Gaussian fit of the whole histogram, with median value $\alpha_{S_\mu} =-0.78$ and standard deviation $\sigma=0.39$. The blue line is the Gaussian fit for $\alpha_{S_\mu}-\sigma\le\alpha_{S}\le\alpha_{S_\mu}+\sigma$. Spectral indices $\alpha_{S}\le\alpha_{S_\mu}-\sigma$ correspond to regions in the 44 GHz image  of low flux density (S/N$<$100), while spectral indices $\alpha_{S}\ge\alpha_{S_\mu}+\sigma$ are associated with regions of the 18 GHz image where S/N$<$100.}
\label{fig:Histo_tot}
\end{center}
\end{figure}

\begin{figure*}[htp]
\begin{center} 
\vspace{-0.0mm}
\advance\leftskip0mm
\includegraphics[width=120mm,angle=0]{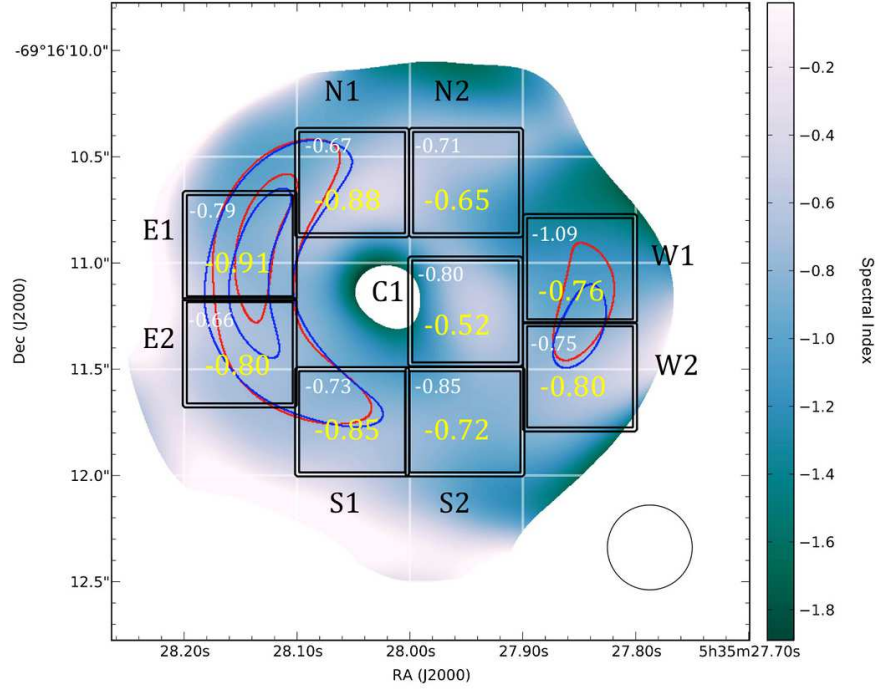}
\caption{Two frequency (18--44 GHz) spectral index intensity map with indication of nine $0\farcs5\times0\farcs5$ regions ({\it black squares}) used for the temperature-temperature (T-T) plots presented in Figure~\ref{fig:T-Tplots}.  For each region, the spectral index values, $\alpha_{TT}$, derived from the T-T plots,  are noted in yellow, while the median values derived from the spectral map, $\alpha_{S_\mu}$, are in white font (see Figure~\ref{fig:T-Tplots}). The contours of the 60\% and 90\% levels of the peak flux density of the 44 and 18 GHz images, resolved with a $0\farcs4$ circular beam, are shown in blue and red, respectively.}
\label{fig:T-Tmap}
\end{center}
\end{figure*}

\begin{figure*}[htp]
\begin{center} 
\vspace{-10.0mm}
\advance\leftskip0mm
\includegraphics[width=160mm,angle=0]{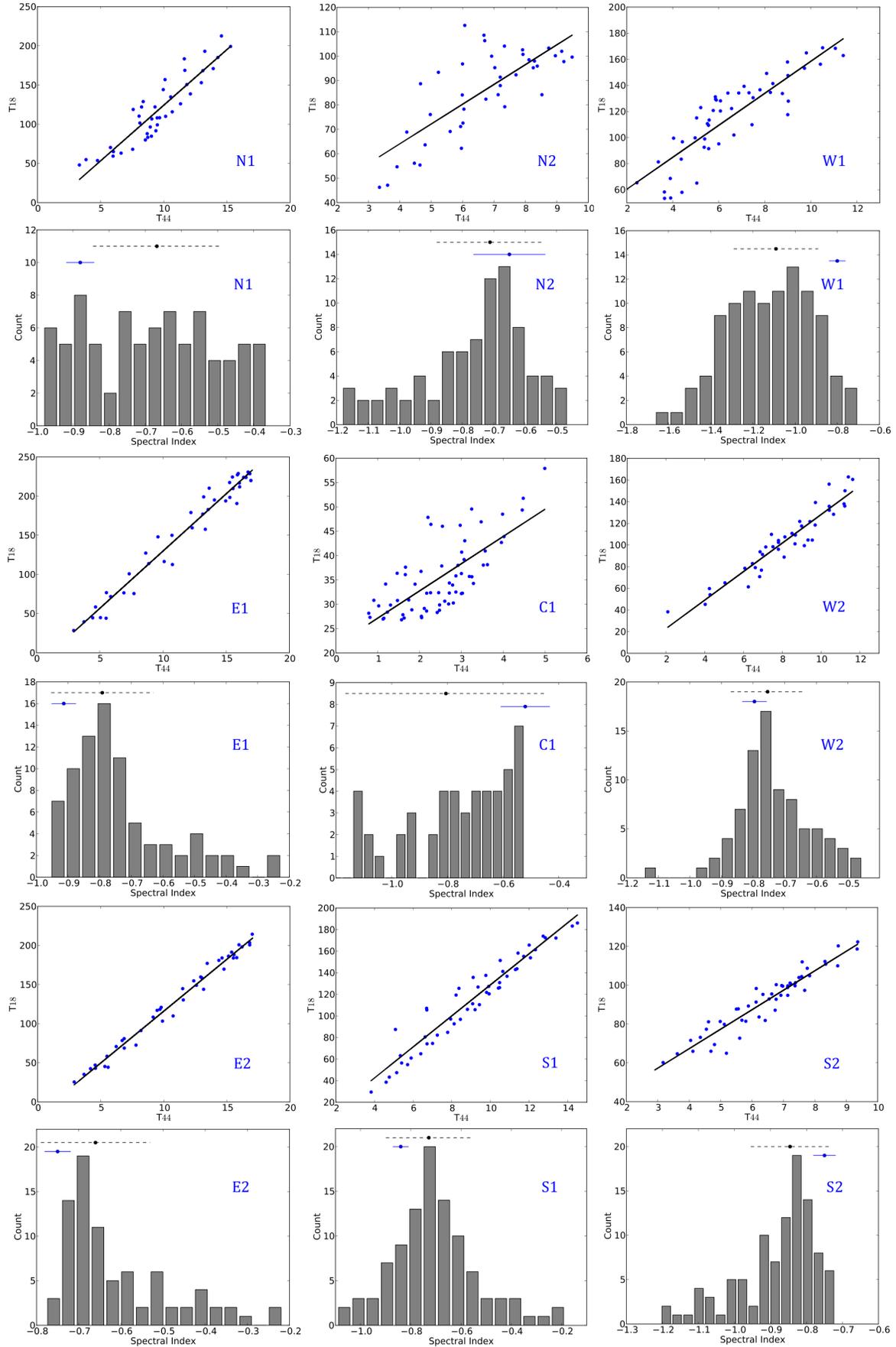}
\caption{Two frequency (18--44 GHz) temperature-temperature (T-T) plots for nine remnant regions, designated as in Figure~\ref{fig:T-Tmap}. For each region, the histogram of the distribution of the spectral index values, $\alpha_{S}$,  is also shown, as obtained from the spectral index intensity map. In each histogram figure,  the solid black circle indicates the related median spectral index and the dashed black line corresponds to the standard deviation, as derived from the Gaussian fit of the histogram. The solid blue circle indicates the spectral index derived from the linear-regression fit of each T-T plot, $\alpha_{TT}$, and the blue bar indicates the $1-\sigma$ error. The values of $\alpha_{TT}$ and $\alpha_{S_\mu}$ associated with each region are listed in Table~\ref{tab_alpha}.}
\label{fig:T-Tplots}
\end{center}
\end{figure*}

\begin{figure}[!tp]
\begin{center}
\vspace{-0.0mm}
\advance\leftskip-4mm
\includegraphics[width=95mm,angle=0]{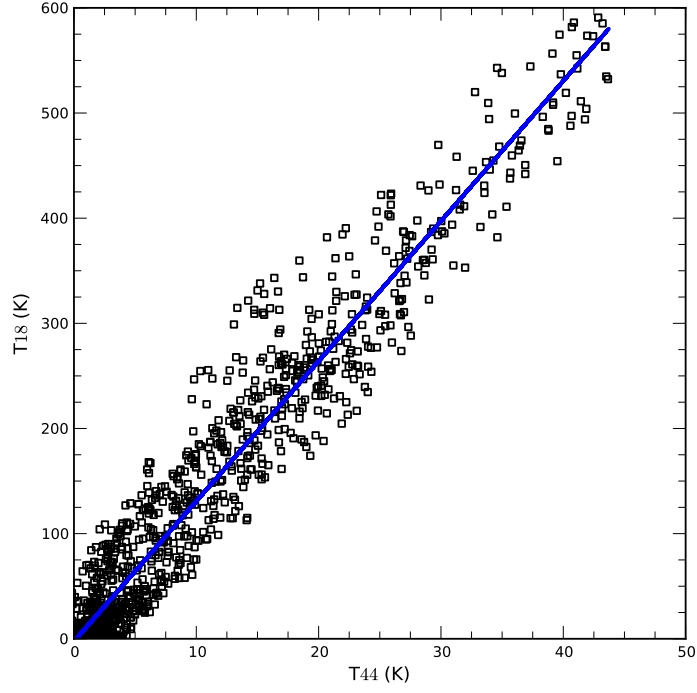}
\caption{Temperature-temperature (T-T) plot of the whole SNR at 44 GHz and 18 GHz ({\it black squares}). The best fitting straight line corresponds to $\alpha=-0.80\pm0.01(\pm0.05)$ ($S=\nu^{\alpha}$), where the two errors represent random and absolute flux density scale errors, respectively.}
\label{fig:T-T_tot}
\end{center}
\end{figure}

\begin{figure*}[htb]
\begin{minipage}[c]{88mm}
\begin{center}
\vspace{-0.5mm}
\advance\leftskip-4.5mm
\includegraphics[width=88.0mm, angle=0]{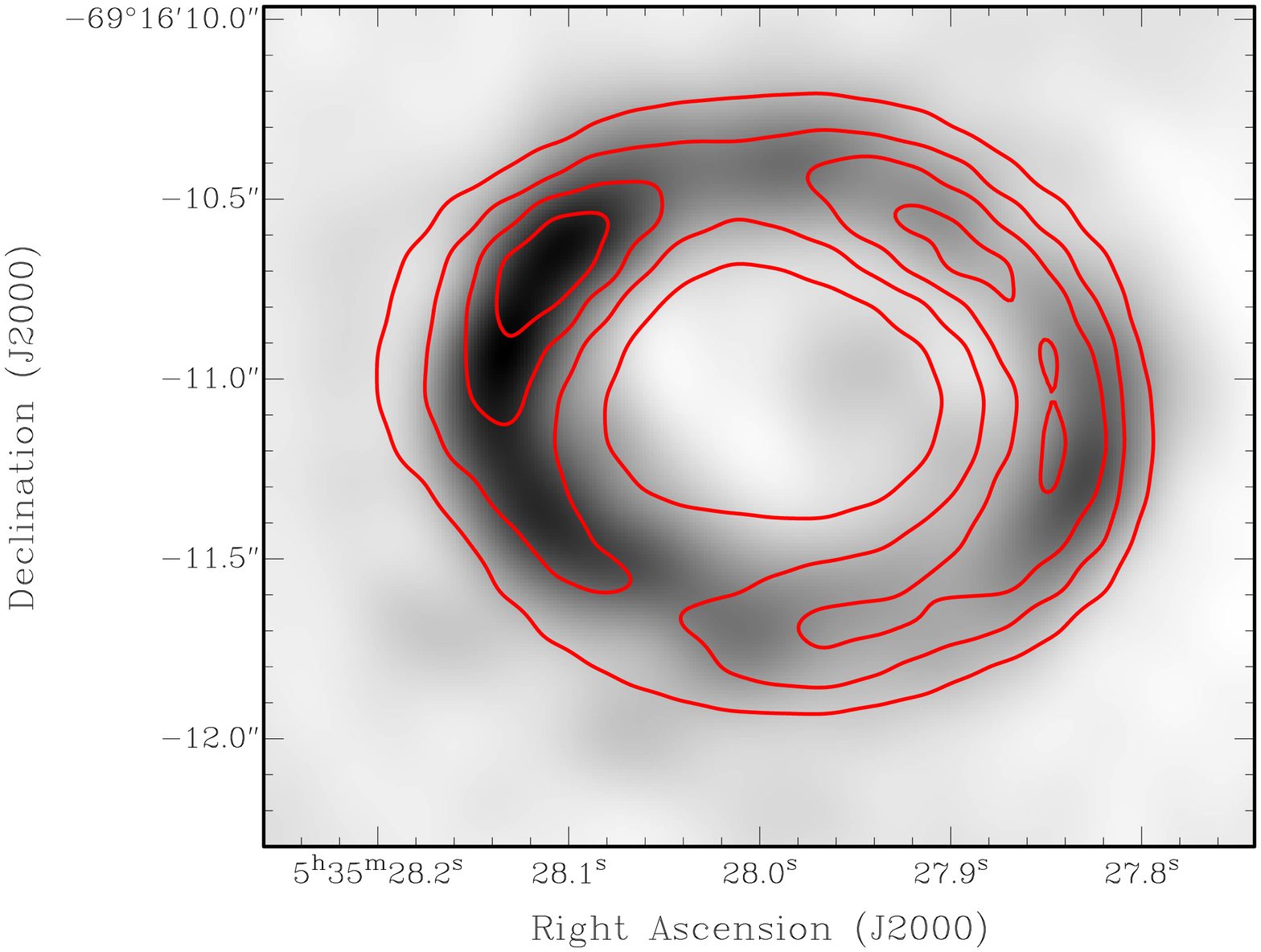}
\end{center}
\end{minipage}
\begin{minipage}[c]{88mm}
\begin{center}
\vspace{+0.5mm}
\advance\leftskip-2.0mm
\includegraphics[width=94.0mm, angle=0]{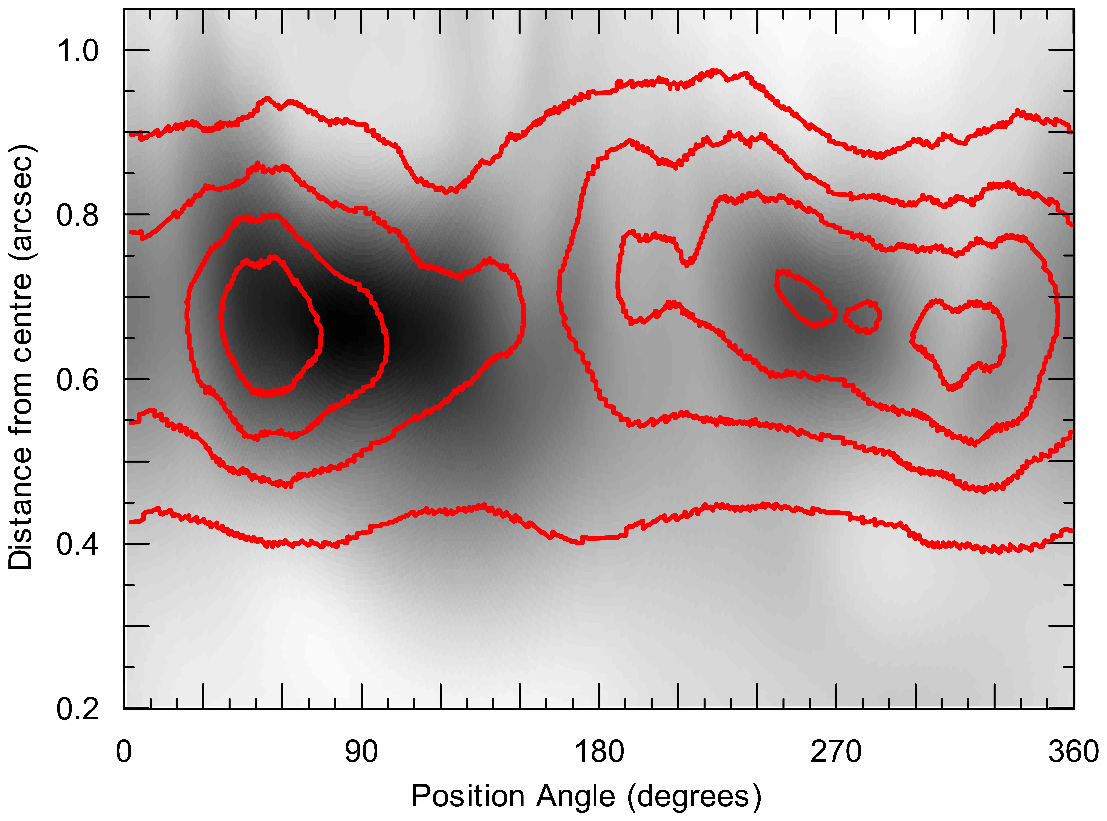}
\end{center}
\end{minipage}
\caption{{\it Left}: Overlay of the {\it Chandra} 0.5--8 keV image ({\it contours}) derived from observations performed in September 2011 \citep{hel12}, on the 44 GHz diffraction-limited image from 2011 ATCA observations. The X-ray contours are at levels 20\%--80\% of  the emission, in steps of 20\%. {\it Right}: We show the 44 GHz diffraction-limited image converted to polar coordinates to visualise asymmetries in the 2--dimensional radial distribution. In this image, the position angle and the projected radial distance from the geometrical centre of the remnant, are the new coordinates. The emission intensity is conserved in the conversion.}
\label{fig:44_X-ray_polar_PA}
\end{figure*}

\begin{figure}[!htp]
\begin{center}
\vspace{2.0mm}
\includegraphics[width=90mm,angle=0]{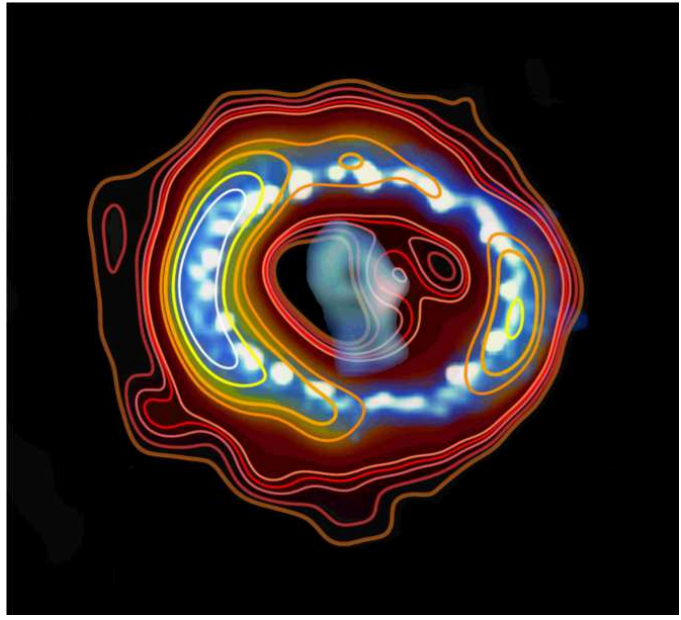}
\caption{Overlay of the 44 GHz image produced from ATCA observations performed in 2011 January and November ({\it brown--yellow} colour scale for shades and contours) on the {\it HST} image derived from 2011 observations\protect\footnotemark[2] ({\it blue--white} colour scale).}
\label{fig:44+Hbl}
\end{center}
\end{figure}

\begin{figure}[htp]
\begin{center}
\vspace{0.0mm}
\advance\leftskip1.0mm
\includegraphics[trim=0mm 0.0mm 3.0mm 0mm, clip=true,width=90mm,angle=0]{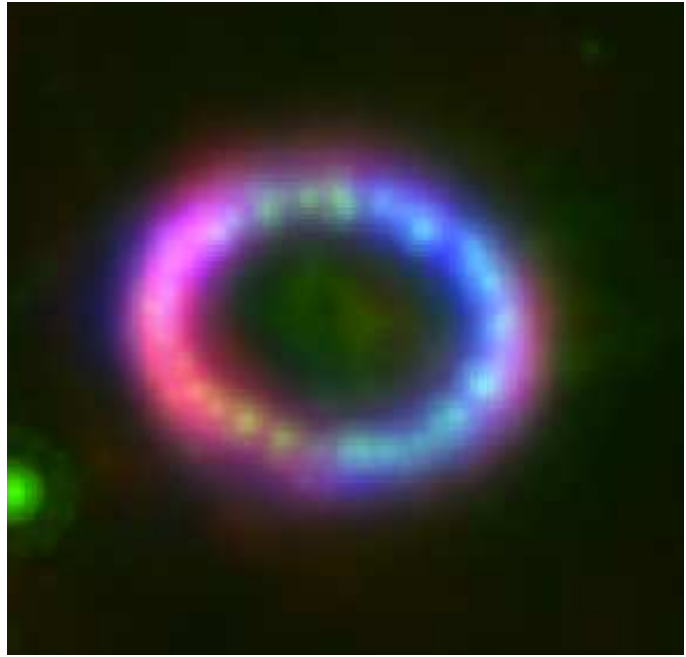}
\caption{RGB overlay of the optical image from  2011 {\it HST} observations\protect\footnotemark[2] ({\it green}), the X-ray image from {\it Chandra} observations performed in 2011 \citep{hel12} ({\it blue}) and the 44 GHz image from ATCA observations in 2011 January and November ({\it red}).}
\label{fig:RGB:44+Hbl+X-ray2}
\end{center}
\end{figure}

\begin{figure}[!htp]
\begin{center}
\vspace{0.5mm}
\advance\leftskip-4.5mm
\includegraphics[width=97mm,angle=0]{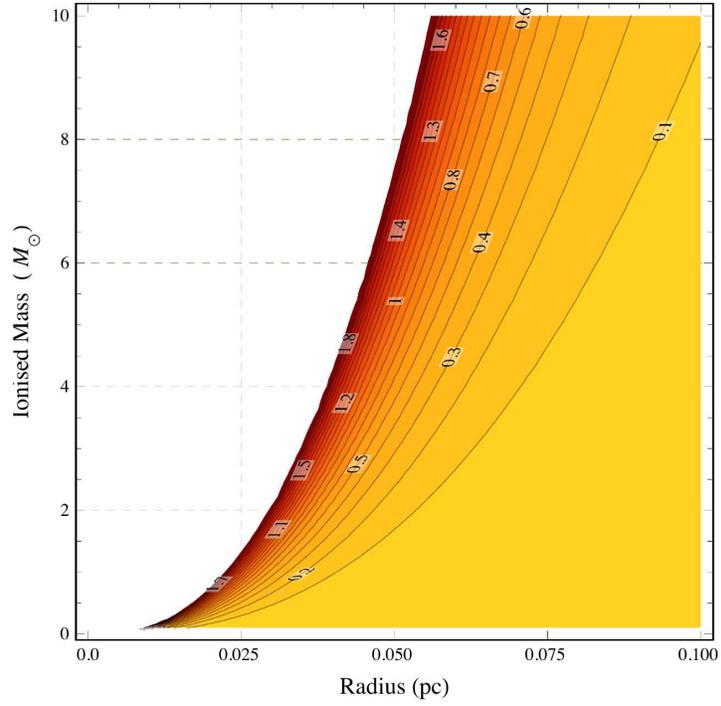} 
\caption{Variation of the optical depth, $\tau_{ff}$, of an idealised H{\sc ii} region in  SNR 1987A, approximately located at the ejecta site. The ionised mass in the region is assumed to vary in the range $0 < M_{ej}\lesssim 10 M_{\sun}$, and to be uniformly distributed within a spherical volume of size $0< R \lesssim 0\farcs4$, which corresponds to a path length up to 0.1 pc  along the line-of-sight. The yellow colour  indicates low free-free opacity; the brown colour indicates high opacity.}
\label{fig:ff_absorb}
\end{center}
\end{figure}

\acknowledgments 
We thank Ray Norris for the assistance provided as duty astronomer during the site-based ATCA observations in 2011 January, and Robin Wark for assistance both on site and during the remote observing session in 2011 November. We thank Robert Kirshner and Peter Challis for providing the {\it HST} images and Eveline Helder for providing the {\it Chandra} image. 
GZ also would like to thank Gerhardt Meurer for help with the {\it HST} images, Tobias Westmeier and Attila Popping for useful feedback on radio imaging techniques.  Figure~\ref{fig:spectral_18-44_divide} utilizes the cube helix colour scheme introduced by \citet{gre11}. GZ has been partially supported by scholarships from ICRAR and the University of Western Australia. Parts of this research were conducted by  CAASTRO, through project number CE110001020. The Australia Telescope Compact Array is part of the Australia Telescope, which is funded by the Commonwealth of Australia for operation as a National Facility managed by CSIRO.


\clearpage


%
%
  
\clearpage

%
%
  
\clearpage

%
%

\clearpage

%
%

\clearpage

%
%

\clearpage

%
%

\clearpage

%
%

\clearpage

%
%

\clearpage

%
%

\clearpage

%
%

\clearpage

%
%

\clearpage

%
%

\clearpage
\bigskip

%
%

\clearpage

%
%

\clearpage

%
%

\clearpage

%
%

\clearpage

%
%

\vspace{10.0mm}

%
%

\clearpage

%
%

\end{document}